\documentclass{elsart}

\usepackage[square,comma]{natbib}

\usepackage{graphics}

\usepackage{amssymb}

\newcommand{\la}{\langle}
\newcommand{\ra}{\rangle}

\newcommand{\hphi}{{\hat \phi}}
\newcommand{\hpi}{{\hat \pi}}

\newcommand{\hrho}{{\hat \rho}}
\newcommand{\tphi}{{\tilde \phi}}
\newcommand{\tpi}{{\tilde \pi}}
\newcommand \bp{{\mathbf p}}
\newcommand \bk{{\mathbf k}}

\newcommand \bq{{\mathbf q}}
\newcommand \br{{\mathbf r}}

\newcommand \bv{{\mathbf v}}

\newcommand{\cV}{{\cal V}}

\newcommand{\cO}{{\cal O}}

\newcommand{\DPi}{{\Delta \Pi}}
\newcommand{\bW}{\mbox{\boldmath $W$}}
\newcommand{\bM}{\mbox{\boldmath $M$}}

\newcommand \Ep{{\omega_\bp}}

\begin{document}

\begin{frontmatter}


\rightline{UT-Komaba/08-03v4}

\title{Quantized meson fields in and out of equilibrium. I : 
Kinetics of meson condensate and quasi-particle excitations}


\author{T. Matsui and M. Matsuo}

\address{Institute of Physics, University of Tokyo \\
Komaba, Tokyo 153-8902, Japan}

\begin{abstract}
We formulate a kinetic theory of self-interacting meson fields
with an aim to describe the freezeout stage of the space-time evolution
of matter in ultrarelativistic nuclear collisions.  Kinetic equations are 
obtained from the Heisenberg equation of motion for a single component
real scalar quantum field taking the mean field approximation for 
the non-linear interaction. 
The mesonic mean field obeys the classical non-linear Klein-Gordon 
equation with a modification due to the coupling to mesonic 
quasi-particle excitations which are expressed in terms of the 
Wigner functions of the quantum 
fluctuations of the meson field, namely the statistical average of the 
bilinear forms of the meson creation and annihilation operators. 
In the long wavelength limit, the equations of motion of the diagonal 
components of the Wigner functions take a form of Vlasov equation 
with a particle source and sink which arises due to the non-vanishing
off-diagonal components of the Wigner function expressing coherent 
pair-creation and pair-annihilation process in the presence of 
non-uniform condensate.  We show that in the static homogeneous system, 
these kinetic equations reduce to the well-known gap equation in the Hartree 
approximation, and hence they may be considered as a generalization of the
Hartree approximation method to non-equilibrium systems.  
As an application of these kinetic equations,
we compute the dispersion relations of the collective mesonic excitations 
in the system near equilibrium.

\end{abstract}


\end{frontmatter}



\section{Introduction}

Theoretical study of the space-time evolution of matter in high energy nuclear 
collisions has a long history since the early pioneering works of Fermi and Landau 
employing thermodynamics and fluid mechanics~\citep{Fer50}. 
More recent works~\cite{Bjo83} have been motivated by the prospect of studying 
a new form of matter experimentally by means of very energetic collisions of
heavy nuclei as those currently underway at Brookhaven with Relativistic 
Heavy-Ion Collider (RHIC) and planned at CERN with Large Hadron Collider (LHC). 
The data taken from the RHIC experiments has shown existence of 
a strong anisotropic (elliptic) collective flow of hadrons in non-central 
collisions~\cite{RHICWP05}, 
indicating early local equilibration of dense matter produced by the collision~\cite{Flow}.
This supports the hydrodynamic picture of matter evolution and gives us 
a hope to learn information of the equation of state of dense matter from 
systematic analyses of the data.
 
Much attention has been paid to the early stage of matter evolution where 
a dense plasma of unconfined quarks and gluons has been expected to be 
formed through some complex non-equilibrium processes~\cite{Bay84} .
One anticipates also a breakdown of the hydrodynamic behavior of 
the system when it is diluted sufficiently and the collision time exceeds the 
characteristic time scale of expansion.
This stage of the matter evolution is usually refered to as the freezeout stage.   
The aim of this work is to present a kinetic theory which is designed to 
describe the freezeout stage of the expanding hadron gas. 

The freezeout of the expanding hadronic matter would proceed in several steps;
chemical freezeout of the relative abandunce of hadron spieces may occur
before the kinetic freezeout of the momentum distribution of hadrons. 
Observed relative abandunce of hadrons fit very nicely to a simple picture
that the chemical freezeout occurs at a certain temperature and baryon 
chemical potential~\cite{BS99,RHICWP05}.
Some of the other observables, such as two particle momentum correlations, 
the pion analogue of the Hanbury Brown and Twiss two photon intensity 
interferometry, may be very sensitive to the dynamics of the final kinetic 
freezeout as indicated by the ``HBT puzzle" found in the recent data analysis~\cite{HBT}.

If the initial state of the dense matter formed in the collision is a plasma
of deconfined quarks and gluons, freezeout of the color degrees of freedom
should proceed these processes when the quark-gluon plasma hadronizes. 
Chiral symmmetry~\cite{NJL61}, an approximate global symmetry of QCD which 
becomes exact in the limit of vanishing quark masses and is considered to be 
broken spontaneously in the QCD vacuum, may also play an important
role in the freezeout dynamics.

As a dense hadronic matter formed by ultrarelativistic nuclear collision 
is diluted by the expansion, one expects that the system undergoes a phase 
transition associated with the spontaneous breakdown of the chiral symmetry 
which is restored temporalily after the collision by the formation of a 
quark-gluon plasma.   As the quark-gluon plasma 
hadronizes and the system turns into the confining phase, the system would 
gradually develop a vacuum chiral condensate and remaining 
excitaions would expand in the influence of the growing chiral condensate.
This physical picture has been elaborated in terms of a classical equation of
motion for the chiral condensate; effects of excitations were described in terms
of statistical fluctuations in the classical fields~\cite{Bjo87,RW93}.
The fluctuating condensate described by classical pion field has been termed
Disoriented Chiral Condensate (DCC), emphasizing the symmetry aspect of
the problem~\cite{RW93}.
What was missing in these classical treatments of the meson fields is the existence 
of particle excitations in addition to the condensate.   Inclusion of  particle 
excitations requires the quantization of the fields. 
The effect of the quantum fluctuations of meson fields has been studied 
by Tsue, Vautherin and one of the present authors (TM) \cite{TVM99}
by the functional Schr\"odinger picture formalism \cite{KV89,EJP88}.

The physical picture we have described is very similar to what happens when 
the dilute gas of magnetically trapped atoms of alkali metals cools down by 
evapolation \cite{PS02}.
Some of the atoms condense into the lowest single particle 
level in the trapping external potential forming a Bose-Einstein condensate.   
The dynamics of such a system may be described by the coupled equations 
of motion for the condensate, or the Gross-Pitaevskii equation
\cite{GP61}
, and the kinetic equation, 
or the Boltzmann-Vlasov equations, for the phase space distribution of 
the excitations in the presence of the condensate~\cite{ZNG99,ITG99}.

In this paper we will show that a similar set of equations can be derived 
for a system of interacting mesons described by the relativistic quantum 
field theory by the mean field approximation.   This approximation 
corresponds to a neglect of all correlations in the system~\cite{KB62}.  
We will make no attempt in this work to justify this approximation and 
leave it as an open problem for further study. 

In this context we note that field theoretical derivation of a Boltzmann-type 
kinetic equation has been given by many others~\cite{CH88}.  
Most of these works focuses on the derivation of the collision terms.
The present work is distinguished from these works in the emphasis of 
the role of the mean field in the evolution of the system coupled with the 
quasi-particle excitations in the same spirit as in \cite{EJP88,TVM99}.
We omit the effect of the quasi-particle collisions in this work.  This procedure
may be reasonable, at least as a first step, for describing dynamical aspect of 
the freeze-out process: even in the absence of collisions, interactions between 
the quasi-particles and the evolving condensate would affect the final 
particle distribution. 
But the present approach is not adequate, however, for the early thermalization 
problem, where the collision terms play essential role~\cite{Bay84}.
\footnote{We note that the mean fields may also play important role 
in the thermalization problem, for example via the non-Abelian analogue 
of the plasma instability \cite{ALMY05}, or by the anomalous enhancement of 
collision rates in a turbulent plasma evolution\cite{ABM06}.}
We adopt the standard Heisenberg picture instead of the functional 
Schr\"odinger picture used in \cite{EJP88,TVM99} since we found it more 
straightforward to see the connection to familiar semi-classical kinetic equation.  
In this first of a series of papers we shall use one-component real scalar field 
model interacting via a $\phi^4$ self-interaction term in order to concentrate 
on the presentation of the basic features of the theory.  Analysis with 
multi-component scalar model with $O(N)$ symmetry will be deferred for 
the forthcoming paper~\cite{MM2}.  

In the next section we formulate a quantum kinetic theory for quantum scalar 
meson field, starting from the Heisenberg equations for quantum field of
one-component real scalar.    The mean field approximation replaces the 
products of quantum fields by 
the products of classical fields and the statistical average of bilinear forms of 
the quantum fluctuations.    The latter is expressed in terms of the Wigner
functions which is reduced to the single particle distribution function in
the classical Boltzmann equation.   Our theory contains another forms of 
the Wigner functions which have no classical counter parts and arises due 
to the coherent pair creation and annihilation processes in non-uniform
systems.   Only in uniform systems, these {\it off-diagonal} 
components of the fluctuation can be eliminated by suitable redefinition 
of the particle mass.    Appearance of the off-diagonal Wigner functions
is reminiscent of the anomalous propagators in the microscopic theory 
of superconductivity~\cite{FW71,Gor58,Nam60}.  Similar structure also 
appears in the theory of Bose-Einstein Condensate \cite{ITG99}. 
Some details of the mean field calculation is given in Appendix.

In section 3, we apply our method to uniform systems and show that each
mode characterized by the particle momentum obeys non-linear 
forced oscillatory motion.  In section 4 we show that in equilibrium these 
kinetic equations are reduced to a gap equation~\cite{DJ74} which 
determine the equilibrium amplitude of the condensate and the mass 
parameter as a function of the temperature.   In this paper, we ignore
the effect of the divergent vacuum polarizations which requires a subtle
renormalization procedure in the mean field approximation~\cite{BG77}.
The solution exhibits characteristic features of the first order phase transition. 

In section 5, we study slowly varying non-uniform systems.  Taking 
long wavelength approximation, the equations of motion of the
diagonal components of the Wigner functions are reduced to a Vlasov
equation in a form generalized by Landau for quasi-particles excitations 
in quantum Fermi liquid~\cite{Lan57,BP91} with the quasi-particle energy given 
in the mean field approximation.  
Non-vanishing off-diagonal components of the Wigner functions 
generate extra terms in the Vlasov equation which may be interpreted 
as particle source and sink terms.   

In section 6, we compute the dispersion relations of excitations
of the system near equilibrium.  Solving the coupled kinetic equations by
linearizing the equations with respect to small oscillatory deviations from 
equilibrium solution we obtain dispersion relation of the excitation
modes in the system near equilibrium.
We find the continuum of quasi-particle excitations in the entire space-like 
energy-momentum region in addition to the continuum in time-like region
due to the (thermally induced) pair creation.  
We found that in the low temperature phase the meson pole shifts due to 
the coupling to the quasi-particle continua.
The effective meson mass vanishes at the edge of the spinodal instability 
line of the first order transition.\footnote{Softening of the $\sigma$ meson 
mode has been studied first by Hatsuda and Kunihiro as a precursory 
phenomenon of the chiral phase transition using the Nambu-Jona-Lasinio 
model with quark fields.\cite{HK85}}
We also found that the coupling of the 
off-diagonal components of the Wigner function plays an important role 
to prevent appearance of undamped tachyonic sound mode which propagates 
with a velocity greater than that of light.    

A short summary of the paper is given in section 7 with remarks on
the remaining problems. 

\section{Kinetic equations for the meson condensate and quasi-particle 
excitations}

In this section we derive quantum kinetic equations which describes the time 
evolution of the meson condensate coupled with mesonic quasi-particle 
excitations. We use the natural unit $\hbar = c = 1$ throughout this paper.

\subsection{Quantized real scalar field in the Heisenberg representation}
We first take a simple model of a self-interacting real scalar field in the
Heisenberg picture.  The Hamiltonian is given by 
\begin{equation}
H = \int d \br \left[ \frac{1}{2} \hpi^2 + \frac{1}{2} (\nabla \hphi )^2 + \cV [ \hphi ] \right]
\end{equation}
where
\begin{equation}
\cV [ \hphi ] = \frac{1}{2} m^2 \hphi^2 + \frac{\lambda}{4!} \hphi^4 .
\end{equation}
Here the scalar field $\hphi$ and its canonical conjugate momentum field 
$\hpi$ are quantized by the equal-time commutation relations:
\begin{eqnarray}
[ \hphi ( \br, t ), \hpi (\br', t ) ]  & = &   i \delta ( \br - \br' ) \\ 
\left[ \hphi ( \br, t ), \hphi (\br', t ) \right] & = & [\hpi ( \br, t ), \hpi ( \br', t ) ] = 0  
\end{eqnarray}
The Heisenberg equation of motion of the quantum field $\hphi (\br, t )$ is 
given by
\begin{equation}
\frac{\partial \hphi}{\partial t}  = - i [ \hphi, H ] =  \hpi ( \br, t )
\end{equation}
while the equation of motion of the canonical conjugate field $\hpi (\br, t )$ 
becomes 
\begin{equation}
\frac{\partial \hpi}{\partial t}  = - i [ \hpi, H ] = ( \nabla^2 - m^2 ) \hphi (\br, t ) 
- \frac{1}{3!} \lambda \hphi^3 (\br, t )
\end{equation}
Elimination of the field momentum $\hpi$ from these equations yields a
modified Klein-Gordon equation for the quantum scalar field $\hphi$:
\begin{equation}\label{KG}
\Box \hphi  (\br, t ) + m^2 \hphi  (\br, t ) = - \frac{1}{3!} \lambda \hphi^3 (\br, t )
\end{equation}
where $\Box = \partial^2/\partial t ^2 - \nabla^2$.

\subsection{Density matrix and Gaussian Ansatz}
We are interested in the time evolution of the system described by the density operator
\begin{equation}
\hrho = \sum_s | \Psi_s \ra p_s \la \Psi_s | 
\end{equation}
where $\{ | \Psi_s \ra \}$ are a set of normalized wave functions 
and $p_s$ is the probability distribution for a mixed state described by this
density matrix so that it satisfies
\begin{equation}
\sum_s p_s = 1
\end{equation}
This density matrix can be also expressed in terms of some complete
set ${ | \alpha \ra }$ of the wave functions of our Hilbert-Fock space as 
\begin{equation}
\hrho = \sum_{\alpha, \beta} | \alpha \ra \rho_{\alpha \beta} \la \beta |
\end{equation}
where 
\begin{equation}
\rho_{\alpha \beta} = \sum_s \la \alpha | \Psi_s \ra  p_s \la \Psi_s | \beta \ra
\end{equation}
  
All physical information of the system to be described are contained 
in a specific form of the density matrix. 
In thermodynamic equilibrium, the density matrix is given by
\begin{equation}
\hrho_{\rm eq} = Z^{-1} e^{ - \beta H} 
\end{equation}
with 
\begin{equation}
Z = {\rm tr} \left[ e^{ - \beta H} \right] = e^{\beta F (T) }
\end{equation}
where $F (T)$ gives the Helmholtz free energy of the system at temperature 
$T = 1/\beta$.

In the Heisenberg picture the density matrix is time-independent since the wave 
functions are time-independent; all time dependence arises from the time 
dependence of an operator:
\begin{equation}
\la {\hat \cO} (t) \ra = {\rm tr} \left[ {\hat \cO} (t) \hrho \right]
\end{equation}
For example, we define the classical condensate fields by the statistical average
of the quantum fields,
\begin{eqnarray}
\phi_c ( \br, t ) & = & \la \hphi (\br , t) \ra, \\
\pi_c ( \br, t ) & = & \la \hpi (\br , t) \ra
\end{eqnarray}
In the following we take a Gaussian Ansatz for the density matrix:
\begin{equation}\label{gaussian1}
\la \tphi^n ( \br, t ) \ra = 0 \qquad  \mbox{for odd integer}; n 
\end{equation}
and 
\begin{equation}
\la \tphi^n ( \br, t ) \ra = \frac{n!}{m!2^m} \la \tphi^2 ( \br, t ) \ra^m  
\qquad \mbox{for even integer}; n = 2 m
\end{equation}
where $\tphi (\br, t)$ and $\tpi (\br, t) $ are shifted field operators defined by 
\begin{eqnarray}
\tphi ( \br, t) & = & \hphi (\br , t ) - \phi_c (\br , t) \label{phi-a} \\
\tpi ( \br, t) & = & \hpi (\br , t ) - \pi_c (\br , t)  \label{pi-a} 
\end{eqnarray}
The shifted field operators obey the same equal-time commutation relations as the 
original fields:
\begin{eqnarray}
[ \tphi ( \br, t ), \tpi (\br', t ) ]  & = &   i \delta ( \br - \br' ) \label{q1} \\
\left[ \tphi ( \br, t ), \tphi (\br', t ) \right] & = & [\tpi ( \br, t ), \tpi ( \br', t ) ] = 0  \label{q2} 
\end{eqnarray}

We will show that this choice of the density matrix is a non-equilibrium generalization 
of the Hartree approximation in equilibrium. It has been alternatively introduced in 
the functional Schr\"odinger representation~\cite{EJP88,TVM99}


\subsection{The Wigner functions}

We introduce the one-particle Wigner function by
\begin{equation}
F (\bp, \bk, t ) = \la a_{\bp + \bk/2}^\dagger (t) a_{\bp - \bk/2} (t)  \ra
\end{equation}
Here the particle creation and annihilation operators may be defined
in terms of the Fourier transforms of the shifted fields
\begin{eqnarray}
\tphi_\bp (t) & = &  \int d \br e^{- i \bp \cdot \br} \tphi (\br, t), \qquad
\tpi_\bp (t)  =  \int d \br e^{ - i \bp \cdot \br} \tpi (\br, t) 
\end{eqnarray}
as
\begin{eqnarray}
a_\bp (t) & = & \frac{1}{\sqrt{2 \Ep}}   
\left[ \Ep \tphi_\bp (t)  + i \tpi_{- \bp} (t) \right] \\
a^\dagger_\bp (t) & = & \frac{1}{\sqrt{2 \Ep}}   
\left[ \Ep \tphi_\bp (t)  -  i \tpi_{- \bp} (t) \right] 
\end{eqnarray}
with
\begin{equation}
\Ep = \sqrt{ p^2 + \mu^2}
\end{equation}
These relations are rewritten as:
\begin{eqnarray}
\tphi (\br, t) & = & \sum_\bp e^{i \bp \cdot \br} \frac{1}{\sqrt{2 \Ep}} 
\left[ a_\bp (t) + a_{-\bp}^\dagger  (t) \right] \\
\tpi (\br, t) & = & i \sum_\bp e^{i \bp \cdot \br} \sqrt{\frac{\Ep}{2}} 
\left[ a_\bp^\dagger (t) - a_{-\bp}  (t) \right]
\end{eqnarray}
The quantization rules (\ref{q1}, \ref{q2}) are transcribed to 
\begin{eqnarray}
[ a_\bp (t) , a_{\bp'}^\dagger (t) ] & = & \delta_{\bp, \bp'} \\
\left[ a_\bp (t) , a_{\bp'} \right (t) ] & = & 
[ a_\bp^\dagger (t) , a_{\bp'}^\dagger (t) ] = 0 .
\end{eqnarray}
with which we may interpret $a_\bp$ ($a^\dagger_\bp$) as annihilation (creation)
operator of ``particle excitation" with momentum $\bp$.

We also introduce the following other forms of the Wigner functions:
\begin{eqnarray}
G ( \bp, \bk, t ) & = & \la a_{- \bp - \bk/2} (t) a_{\bp - \bk/2} (t)  \ra, \\
{\bar G} ( \bp, \bk, t ) & = & \la a_{\bp + \bk/2}^\dagger (t) a_{- \bp + \bk/2}^\dagger  (t)  \ra, \\
{\bar F} ( \bp, \bk, t ) & = & \la a_{-\bp - \bk/2} (t) a_{- \bp + \bk/2}^\dagger  (t)  \ra,
\end{eqnarray}
These Wigner functions are not independent but are related to each other.
The complex conjugate of the Wigner functions are given by 
\begin{eqnarray}
F^* ( \bp, \bk, t ) & = & F ( \bp, - \bk, t ), \label{Fstar}\\
{\bar F}^* ( \bp, \bk, t ) & = & {\bar F} (  \bp, - \bk, t ) \label{Fbarstar} \\
{\bar G}  ( \bp, \bk, t )  & = &  G^* (  \bp, - \bk, t ) \label{Gstar}
\end{eqnarray}
where the asterisk ($^*$) stands for the complex conjugate.
The commutation relations imply also that $G ( \bp, \bk, t ) $ and ${\bar G} ( \bp, \bk, t ) $ 
are even functions of $\bp$ 
\begin{eqnarray}
G  ( \bp, \bk, t )  & = &  G ( - \bp, \bk, t )  \\
{\bar G}  ( \bp, \bk, t )  & = &  {\bar G} ( - \bp, \bk, t )  
\end{eqnarray}
and 
\begin{equation}
{\bar F} ( \bp, \bk, t ) =  F ( - \bp, \bk, t ) + \delta_{\bk, 0 }
\end{equation}
The four  Wigner functions may be grouped together to form a matrix form of the 
Wigner function
\begin{equation}
\bW (\bp, \bk, t ) =
\left(
\begin{array}{cc}
F ( \bp, \bk, t)   &    {\bar G} ( \bp, \bk, t )  \\
G  ( \bp, \bk, t )  &  {\bar F}  ( \bp, \bk, t )
\end{array}
\right)
\end{equation}

The appearance of the ``off-diagonal" components of the Wigner functions
is reminiscent of the anomalous propagators in the BCS theory of superconductivity
which arises due to the presence of fermion pair condensate~\cite{FW71,Gor58,Nam60}. 
Note that our definition of the matrix components of the Wigner function is 
slightly different from those for the propagators.

We write the Fourier transforms of the Wigner functions as 
\begin{eqnarray}
f ( \bp, \br, t ) & = & \sum_\bk e^{- i \bk \cdot \br } F ( \bp, \bk, t ),  ~~~
{\bar f} ( \bp, \br, t )  =  \sum_\bk e^{- i \bk \cdot \br } {\bar F} ( \bp, \bk, t ),   
\label{fourierf} \\ 
g( \bp, \br, t ) & = & \sum_\bk e^{- i \bk \cdot \br } G ( \bp, \bk, t ),  ~~~
{\bar g} ( \bp, \br, t )  =  \sum_\bk e^{- i \bk \cdot \br } {\bar G} ( \bp, \bk, t )
\label{fourierg} 
\end{eqnarray}
The (\ref{Fstar}) and (\ref{Fbarstar}) imply that $f (\bp, \br, t )$
and ${\bar f} (\bp, \br, t ) $ are real functions 
and are related to each other by 
\begin{equation}\label{fbar}
{\bar f} (\bp, \br, t )  =  f (- \bp, \br, t ) + 1,  
\end{equation}
while (\ref{Gstar}) implies that $g (\bp, \br, t )$ and ${\bar g} (\bp, \br, t )$ are 
complex conjugate to each other:
\begin{eqnarray}\label{gbar}
g^* (\bp, \br, t ) = {\bar g} (\bp. \br, t )
\label{gstar-gbar}
\end{eqnarray}

Here we have chosen the particle ``mass"  $\mu$ to be different from the mass parameter $m$ 
in the original Hamiltonian.   Physical particle mass for interacting fields
is generally different from the mass parameter in the Hamiltonian or Lagrangian 
due to the effect of interaction, e.g. renormalization with or without the spontaneous symmetry 
breaking.   It may also depend on the physical conditions described by the statistical average 
with the density matrix $\hrho$.   Since we are interested in non-equilibrium time-evolution of 
the system where the physical particle mass may not have a definite meaning in the intermediate 
states, we consider here $\mu$ just as a parameter to be chosen at our discretion, 
for an appropriate choice of the initial conditions specified by the Gaussian density matrix.
A different choice of this mass parameter would give different definition for ``particle excitations"; 
the Wigner functions thus depend on the particular choice of the mass parameter.
        
Suppose we take a different particle mass $\mu'$ to define the particle creation and annihilation operators 
\begin{eqnarray}
b_\bp (t) & = & \frac{1}{\sqrt{2 \Ep'}}   
\left[ \Ep' \tphi_\bp (t)  + i \tpi_{- \bp} (t) \right] \\
b^\dagger_\bp (t) & = & \frac{1}{\sqrt{2 \Ep'}}   
\left[ \Ep' \tphi_\bp (t)  -  i \tpi_{- \bp} (t) \right] 
\end{eqnarray}
with
\begin{equation}
\Ep' = \sqrt{ p^2 + \mu'^2}
\end{equation}
These new particle creation and annihilation operators should also obey the 
commutation relations,
\begin{eqnarray}
[ b_\bp (t) , b_{\bp'}^\dagger (t) ] & = & \delta_{\bp, \bp'} \\
\left[ b_\bp (t) , b_{\bp'} \right (t) ] & = & 
[ b_\bp^\dagger (t) , b_{\bp'}^\dagger (t) ] = 0 .
\end{eqnarray}
so that they are related to the original ones by the Bogoliubov transformation:
\begin{eqnarray}
b_\bp ~ & = &  \cosh \alpha_p ~ a_\bp +  \sinh \alpha_p ~ a^\dagger_{-\bp} \\
b_{- \bp} ^\dagger & = & \sinh \alpha_p ~ a_{\bp} 
+ \cosh \alpha_p  ~  a_{-\bp}^\dagger
\end{eqnarray}
where the real parameter $\alpha_p$ is determined by requiring that they describe 
the same fields:
\begin{eqnarray}
\tphi (\br, t) & = & \sum_\bp 
\frac{e^{i \bp \cdot \br}}{\sqrt{2 \Ep}} \cdot
\left[ a_\bp (t) + a_{-\bp}^\dagger  (t) \right]
= \sum_\bp  
\frac{e^{i \bp \cdot \br} }{\sqrt{2 \omega'_\bp}} \cdot
\left[ b_\bp (t) + b_{-\bp}^\dagger  (t) \right]
\nonumber \\
\end{eqnarray}
and this gives
\begin{equation}
e^{\alpha_p} = \sqrt{ \frac{\omega'_p}{\omega_p}}
\end{equation}
or 
\begin{equation}
\alpha_p = \frac{1}{2}\log{ \left( \frac{\omega'_p}{\omega_p} \right) } 
= \frac{1}{4}\log{ \left( \frac{p^2 + \mu'^2}{p^2 + \mu^2} \right) }
\end{equation}

The new Wigner functions defined by replacing the creation and annihilation by
the new ones are related to the original Wigner functions by:
\begin{eqnarray}\label{rotation}
\bW' (\bp, \bk, t ) & = &
\left(
\begin{array}{cc}
F' ( \bp, \bk, t)   &    {\bar G'} ( \bp, \bk, t )  \\
G'  ( \bp, \bk, t )  &   {\bar F'}  ( \bp, \bk, t )
\end{array}
\right)
\nonumber \\
& = & \bM ( \bp + \bk/2  ) \bW (\bp, \bk, t) \bM (\bp- \bk/2 ) 
\end{eqnarray}
where 
\begin{equation}
\bM ( \bp ) = \left(
\begin{array}{cc}
\cosh \alpha_{\bp} &   \sinh \alpha_{\bp}  \\
\sinh \alpha_{\bp}  &  \cosh \alpha_{\bp}
\end{array}
\right) 
= e^{\alpha_p \tau_1}
\end{equation}
with 
\begin{equation}
\tau_1 = \left(
\begin{array}{cc}
0 &  1  \\
1  &  0
\end{array}
\right), \quad
\tau_2 = \left(
\begin{array}{cc}
0 &  -i  \\
i  &  0
\end{array}
\right), \quad
\tau_3 = \left(
\begin{array}{cc}
1 &   0 \\
0  &  -1 
\end{array}
\right)
\end{equation}
For a small change of the mass parameter $\mu \to \mu + \delta \mu$
the Wigner function will change by
\begin{eqnarray}
\delta \bW (\bp, \bk, t ) & = &
\tau_1 \bW (\bp, \bk, t ) \delta \alpha_{\bp+ \bk/2} 
+ \bW (\bp, \bk, t ) \tau_1 \delta \alpha_{\bp - \bk/2}  \nonumber \\
& = &
 \left( \frac{\tau_1 \bW (\bp, \bk, t )}{4 ( (\bp+ \bk/2 )^2 + \mu^2)} 
+ \frac{\bW (\bp, \bk, t ) \tau_1 }{4 ( (\bp - \bk/2 )^2 + \mu^2)} 
\right) \mu \delta \mu \nonumber \\
\end{eqnarray}
In particular, 
\begin{eqnarray}\label{dF}
\delta F (\bp, \bk, t ) & = & 
\left( \frac{ {\bar G} (\bp, \bk, t ) )}{4 ( (\bp+ \bk/2 )^2 + \mu^2)} +
\frac{G (\bp, \bk, t ) }{4 ( (\bp - \bk/2 )^2 + \mu^2)} \right) \mu \delta \mu
\end{eqnarray}

In uniform equilibrium system the mass parameter may be chosen to "diagonalize" the one-body 
mean field Hamiltonian.   As will be shown later, this procedure will lead to the well-known gap 
equation, a self-consistency condition to determine $\mu$.   If the system is slowly changing 
in time, one may still use such procedures adjusted to slowly varying quasi-equiliblium conditions, introducing a time dependent effective mass as a dynamical parameter to describe such adiabatic 
process.  For such calculations, the relation (\ref{dF}) may be used to describe the adiabatic 
change of the Wigner functions by the change of the mass parameter.

In more general non-equilibrium situations as we expect to encounter at the freeze-out stage 
of expanding matter, however, there may be no such appropriate condition to determine 
the mass parameter.  The situation could even be worse:  if the system goes through unstable 
state with respect to small fluctuation of the field, then the adiabatically determined mass 
parameter would become pure imaginary,  reflecting the extremum, instead of the minimum,  
of the effective potential.   In such case we may keep the value of $\mu$ at a certain real
value reflecting initial conditions.  The instability would then show up as appearance of 
a growing solution to our kinetic equations; which would eventually be stabilized by the 
non-linear interaction.  

To extract the physical information, such as the particle distribution in the final asymptotic 
state, we should use the Wigner function defined with the physical particle mass in the vacuum.  
But these asymptotic physical Wigner functions may be calculated from the Wigner functions
with a different choice of the mass parameter by the relation (\ref{rotation}).

\subsection{Equation of motion in the mean field approximation}

The equation of motion of the classical mean field $\phi_c (\br , t)$ is obtained 
by taking the quantum statistical average of the field equation (\ref{KG}) .
With the Gaussian Ansatz for the density matrix, we find
\begin{equation}\label{nlKG}
\Box 
\phi_c  (\br, t ) + m^2  \phi_c  (\br, t ) 
= - \frac{1}{3!} \lambda \left[ \phi_c^3 (\br, t ) + 3 \la \tphi^2 (\br,t) \ra \phi_c (\br, t ) \right]
\end{equation} 
This equation corresponds to the non-linear Sch\"odinger equation
(also called the Gross-Pitaevskii equation \cite{PS02}) in the theory of Bose-Einstein condensates. 
So we may call this equation {\it non-linear Klein-Gordon equation}. 
The non-linearity arises due to the self-interaction of the classical field $\phi_c (\br, t)$ 
(condensate) and also due to the interaction with fluctuations $ \la \tphi^2 (\br,t) \ra$ 
which also depends on $\phi_c (\br, t)$ implicitly. 
The latter  may be interpreted as due to ``particle excitations", since the fluctuation can be expressed
by the Wigner functions as
\begin{eqnarray}
\la \tphi^2 (\br,t) \ra & = &
\sum_{\bp,\bp'} \frac{e^{i ( - \bp + \bp' ) \cdot \br}} {\sqrt{2\omega_\bp}\sqrt{2 \omega_{\bp'}} } 
\la ( a_{- \bp}  + a_{\bp}^\dagger  )  ( a_{\bp'}  + a_{-\bp'}^\dagger  ) \ra  
\nonumber \\
& = &
\sum_{\bp,\bp'} \frac{e^{i ( - \bp + \bp' ) \cdot \br}} {\sqrt{2\omega_\bp}\sqrt{2 \omega_{\bp'}} } 
\left[ F ( \frac{\bp + \bp'}{2}, \bp - \bp', t) + {\bar F} ( \frac{\bp + \bp'}{2}, \bp - \bp',t) \right.
 \nonumber \\
&& \qquad \left. + G ( \frac{\bp + \bp'}{2}, \bp - \bp',t) + {\bar G} ( \frac{\bp + \bp'}{2}, \bp - \bp', t) \right]
\label{phi2}
\end{eqnarray}
The time-evolution of the classical mean field $\phi_c (\br, t) $ is thus coupled with 
the time-evolution of the Wigner functions.

To derive the equation of motion of the Wigner functions, we need to compute the time-derivative
of the bilinear forms of the operators $a_\bp (t)$ and $a_\bp^\dagger (t)$ which in turn requires computation of the commutators of these operators with the hamiltonian. 
We decompose the original hamiltonian as
\begin{equation}
H = H_0 + H_1+ H_2 + H_3 + H_4
\end{equation}
where $H_0 $ is the classical hamiltonian obtained from $H$ by replacing the quantum fields by 
their classical expectation values and $H_i$ contain the $i$-thrth power of 
the quantum fluctuation $\hphi$ (or $a_\bp$ and $a_\bp^\dagger$). 
A straightforward calculation yields
\begin{eqnarray}
H_1 & = & \int d \br \left[ \pi_c  \tpi + \nabla \phi_c \nabla \tphi + m^2 \phi_c \tphi 
+ \frac{\lambda}{3!} \phi_c^3 \tphi \right] \\
H_2 & = & \int d \br \left[ \frac{1}{2} \tpi^2 + \frac{1}{2} (\nabla \tphi )^2 
+ \frac{1}{2} m^2 \tphi^2( \br, t ) + \frac{\lambda}{4} \phi_c^2 ( \br, t ) \tphi^2 ( \br, t )\right] \\
H_3 & = & \frac{\lambda}{3!} \int d \br  \phi_c ( \br, t )\tphi^3 ( \br, t )\\
H_4 & = & \frac{\lambda}{4!} \int d \br  \tphi^4 ( \br, t ) 
\end{eqnarray}

The commutators of bilinear forms of $a_\bp (t)$ and $a_\bp^\dagger (t)$ with  $H_1$ vanish 
and the commutators with $H_3$ would give either a linear term or the third power of the 
fluctuation, both of which may vanish when taking the average with the Gaussian density matrix.   
What remain to be computed are then the commutators with $H_2$ and with $H_4$.  
They will give either the bilinear form of $a_\bp (t)$ and $a_\bp^\dagger$
or the fourth power of the fluctuations.  The Gaussian average of the resultant equations of
motion of the bilinear field operators would give the desired equations of motion of the Wigner 
functions. Details of this computation is given in Appendix A.

The resultant equation of motion of the Wigner functions may be obtained more easily 
by introducing the mean field Hamiltonian defined by 
\begin{eqnarray}\label{Hmf}
H_{\rm mf} & = & \int d \br \left[ \frac{1}{2} \tpi^2 + \frac{1}{2} (\nabla \tphi )^2 + \frac{1}{2} m^2 \tphi^2 + \frac{1}{2} \Pi ( \br, t ) \tphi^2 \right]  \nonumber \\
& = & \int d \br \left[ \frac{1}{2} \tpi^2 + \frac{1}{2} (\nabla \tphi )^2 
+ \frac{1}{2} \mu^2 \tphi^2 + \frac{1}{2} \DPi ( \br, t ) \tphi^2 \right]  
\end{eqnarray}
where 
\begin{equation}\label{Pi}
\Pi ( \br, t ) = \frac{\lambda}{2} \left( \phi_c^2 (\br, t ) + \la \tphi^2 (\br, t ) \ra \right) 
\end{equation}
and 
\begin{equation}
\DPi( \br, t ) = \Pi ( \br, t ) + m^2 - \mu^2
\end{equation}
In the momentum representation this mean field Hamiltonian may be written as
\begin{equation}\label{Hmf'}
H_{\rm mf} = \sum_\bp \Ep a^\dagger_\bp a_\bp 
+ \frac{1}{2} \sum_{\bp, \bq} \DPi_\bq \cdot 
\frac{ ( a_{\bp} + a^\dagger_{-\bp} ) ( a_{-\bp-\bq} + a^\dagger_{\bp + \bq})}
{\sqrt{2\omega_{\bp}} \sqrt{2 \omega_{\bp + \bq}}} 
\end{equation}
where
\begin{equation}\label{self}
\DPi_\bq (t) = \int d \br e^{- i \bq \cdot \br} ( \Pi (\br, t) +  m^2 - \mu^2 ) 
= \Pi_\bq (t) +  (  m^2 - \mu^2 )  \delta_{\bq, 0}
\end{equation}

The commutator of a bilinear operator product of $a_\bp$ and $a_\bp^\dagger$ 
with this mean-field Hamiltonian is given by
\begin{eqnarray}
[ a^\dagger_{\bp_1} a_{\bp_2} , H_{\rm mf} ]  
& = & - ( \omega_{\bp_1} - \omega_{\bp_2} )  a^\dagger_{\bp_1}  a_{\bp_2} 
- \sum_\bq  \DPi_\bq \cdot
\frac{ ( a_{-\bp_1-\bq} +  a_{\bp_1+ \bq}^\dagger ) a_{\bp_2} }
{\sqrt{2 \omega_{\bp_1 + \bq} } \sqrt{2\omega_{\bp_2}}}
\nonumber \\
& &
\qquad  + \sum_\bq \DPi_\bq \cdot
\frac{a_{\bp_1}^\dagger ( a_{\bp_2 - \bq}   + a_{- \bp_2 + \bq}^\dagger ) }
  {\sqrt{2 \omega_{\bp_1} } \sqrt{2\omega_{\bp_2 - \bq}}}
\end{eqnarray} 
We show in Appendix that the quantum statistical average of this 
commutator with the Gaussian density matrix gives precisely the same result 
for the same statistical average of the commutator with the original Hamiltonian:
\begin{equation}
\la [ a^\dagger_{\bp_1} a_{\bp_2} , H_{\rm mf} ]  \ra
= \la [ a^\dagger_{\bp_1} a_{\bp_2} , H]  \ra
\end{equation}
Therefore one can compute the equations of motion of the Wigner functions 
using this effective Hamiltonian, 
\begin{equation}
i \frac{\partial}{\partial t} F ( \bp, \bk, t ) = 
\la [ a^\dagger_{\bp + \bk/2} a_{\bp - \bk/2} , H ] \ra
= \la [ a^\dagger_{\bp + \bk/2} a_{\bp - \bk/2} , H_{\rm mf} ] \ra
\end{equation}
Using this we find,
\begin{eqnarray}
i \frac{\partial}{\partial t} F ( \bp, \bk, t ) & = &  
- ( \omega_{\bp + \bk/2}  - \omega_{\bp - \bk/2})  F ( \bp, \bk, t )
\nonumber \\
& & \quad  - \sum_\bq \DPi_\bq \cdot
\frac{ F ( \bp + \bq/2, \bk + \bq, t ) + G ( \bp + \bq/2, \bk + \bq, t ) }
{\sqrt{2 \omega_{\bp + \bk/2}}\sqrt{2 \omega_{\bp+ \bk/2 + \bq }}}
\nonumber \\
& &
\qquad  + \sum_\bq \DPi_\bq \cdot
\frac{F ( \bp - \bq/2, \bk + \bq, t ) + {\bar G} ( \bp - \bq/2, \bk + \bq, t ) }
{\sqrt{2 \omega_{\bp - \bk/2}}\sqrt{2\omega_{\bp - \bk/2 - \bq }}}
 \nonumber \\
 \label{eomF}
 \end{eqnarray}
Equations of motion of other three Wigner functions $G ( \bp, \bk, t )$, 
 ${\bar G} ( \bp, \bk, t )$,  ${\bar F} ( \bp, \bk, t )$ can be also computed from 
the commution relations of the product operators $a_{\bp_1} a_{\bp_2}$, 
$a^\dagger_{\bp_1} a^\dagger_{\bp_2}$, $a_{\bp_1} a^\dagger_{\bp_2}$, 
with the mean field Hamiltonian $H_{\rm mf}$, respectively:
We obtain
\begin{eqnarray}
i \frac{\partial}{\partial t} G ( \bp, \bk, t ) & = &  
 ( \omega_{\bp + \bk/2}  +  \omega_{\bp - \bk/2})  G ( \bp, \bk, t )
\nonumber \\
& & \quad  + \sum_\bq \DPi_\bq \cdot
\frac{ G ( \bp + \bq/2, \bk + \bq, t ) + F ( \bp + \bq/2, \bk + \bq, t ) }
{\sqrt{2 \omega_{\bp + \bk/2}}\sqrt{2\omega_{\bp+ \bk/2 + \bq }}}
\nonumber \\
& &
\qquad  + \sum_\bq \DPi_\bq \cdot
\frac{G ( \bp - \bq/2, \bk + \bq, t ) + {\bar F} ( \bp - \bq/2, \bk + \bq, t ) }
{\sqrt{2 \omega_{\bp - \bk/2}}\sqrt{2\omega_{\bp - \bk/2 - \bq }}}
 \nonumber \\
\label{eomG}
 \\
  i \frac{\partial}{\partial t} {\bar G} ( \bp, \bk, t ) & = &  
-  ( \omega_{\bp + \bk/2}  +\omega_{\bp - \bk/2})  {\bar G} ( \bp, \bk, t )
\nonumber \\
& & \quad  - \sum_\bq \DPi_\bq \cdot
\frac{ {\bar G} ( \bp + \bq/2, \bk + \bq, t ) +  {\bar F}  ( \bp + \bq/2, \bk + \bq, t ) }
{\sqrt{2 \omega_{\bp + \bk/2}}\sqrt{2 \omega_{\bp+ \bk/2 + \bq }}}
\nonumber \\
& &
\qquad  - \sum_\bq \DPi_\bq \cdot
\frac{{\bar G} ( \bp - \bq/2, \bk + \bq, t ) +F ( \bp - \bq/2, \bk + \bq, t ) }
{\sqrt{2 \omega_{\bp - \bk/2}}\sqrt{2 \omega_{\bp - \bk/2 - \bq }}}
 \nonumber \\
 \label{eomGbar}
 \end{eqnarray}
 and the equation of motion of ${\bar F} (\bp, \bk, t)$ can be obtained from 
 (\ref{eomF}) by the substitution $\bp \to - \bp$: 
 \begin{eqnarray}
  i \frac{\partial}{\partial t} {\bar F} ( \bp, \bk, t ) & = &  
 ( \omega_{\bp + \bk/2}  - \omega_{\bp - \bk/2}) {\bar F} ( \bp, \bk, t )
\nonumber \\
& & \quad  + \sum_\bq \DPi_\bq \cdot
\frac{ {\bar F} ( \bp + \bq/2, \bk + \bq, t ) +  {\bar G}  ( \bp + \bq/2, \bk + \bq, t ) }
{\sqrt{2 \omega_{\bp + \bk/2}}\sqrt{2 \omega_{\bp+ \bk/2 + \bq }}}
\nonumber \\
& &
\qquad  - \sum_\bq \DPi_\bq \cdot
\frac{{\bar F} ( \bp - \bq/2, \bk + \bq, t ) + G ( \bp - \bq/2, \bk + \bq, t ) }
{\sqrt{2 \omega_{\bp - \bk/2}}\sqrt{2 \omega_{\bp - \bk/2 - \bq }}}
 \nonumber \\
 \label{eomFbar}
 \end{eqnarray}
These equations form a closed system of coupled differential equations
with the non-linear Klein-Gordon equation (\ref{nlKG}) which may be 
rewritten as
\begin{equation}\label{nlKG'}
\Box 
\phi_c  (\br, t ) + ( \mu^2  + \DPi (\br, t) )\phi_c  (\br, t ) = - \frac{\lambda}{3} \phi_c^3 (\br, t ) .
\end{equation}

These four equations of motion of the Wigner function may be combined into a single 
matrix form as
\begin{eqnarray}
i \frac{\partial}{\partial t} \bW (\bp, \bk, t ) & = &  
- \omega_{\bp + \bk/2} \tau_3 \bW (\bp, \bk, t )  + \omega_{\bp - \bk/2}  \bW ( \bp, \bk, t ) \tau_3
\nonumber \\
& & \quad  - \sum_\bq \DPi_\bq \cdot
\frac{ \tau_3 (1 + \tau_1) \bW ( \bp + \bq/2, \bk + \bq, t ) }
{\sqrt{2 \omega_{\bp + \bk/2}}\sqrt{2 \omega_{\bp+ \bk/2 + \bq }}}
\nonumber \\
& &
\qquad  + \sum_\bq \DPi_\bq \cdot
\frac{  \bW ( \bp - \bq/2, \bk + \bq, t ) (1 + \tau_1) \tau_3 }
{\sqrt{2 \omega_{\bp - \bk/2}}\sqrt{2\omega_{\bp - \bk/2 - \bq }}}
 \nonumber \\
 \label{eomW}
 \end{eqnarray}

\section{Uniform system}
For a uniform system, we expect that the classical mean field and 
the self-energy become functions only of time: 
\begin{equation}
\phi_c (\br, t ) = \phi_0 ( t ), \qquad \Pi ( \br, t ) = \Pi_0 ( t ) 
\end{equation}
Thus the non-linear Klein-Gordon equation (\ref{nlKG'}) becomes
 \begin{equation}\label{nlKG0}
{\ddot \phi}_0  (\br, t ) + ( \mu^2  + \DPi_0 ( t) )\phi_0  ( t ) = - \frac{\lambda}{3} \phi_0^3 (t ) .
\end{equation}
and the mean-field Hamiltonian is reduced to
\begin{equation}
H_{\rm mf} = \sum \omega_\bp a^\dagger_\bp a_\bp 
+ \DPi_0 (t) \sum_\bp ( a_\bp + a^\dagger_{-\bp} )  ( a_{- \bp} + a^\dagger_\bp ) 
\end{equation}

In this case the Wigner functions contain non-vanishing components only for the
diagonal elements ($\bk = 0$) so that they may be written as
\begin{eqnarray}
F ( \bp, \bk, t ) & = & F_0 ( \bp, t) \delta (\bk)  \\
G ( \bp, \bk, t ) & = & G_0 ( \bp, t ) \delta (\bk)  \\
{\bar G} ( \bp, \bk, t ) & = & {\bar G}_0 ( \bp, t ) \delta (\bk)  \\
{\bar F} ( \bp, \bk, t ) & = & \left( F_0 ( - \bp, t ) + 1 \right) \delta_{\bk, 0}  
\end{eqnarray}
Then the equations of motion of the Wigner functions become
\begin{eqnarray}
i \frac{\partial}{\partial t} F_0 ( \bp, t )  & = &   
\frac{\DPi_0} {2 \Ep}  \left( {\bar G}_0 ( \bp, t ) - G_0 ( \bp, t ) \right) \\
i \frac{\partial}{\partial t} {\bar F}_0 ( \bp, t ) & = &  
\frac{\DPi_0} {2 \Ep}  \left( {\bar G}_0 ( \bp, t ) - G_0 ( \bp, t ) \right) \\
i \frac{\partial}{\partial t} G_0 ( \bp, t ) & = & 
\left( 2 \Ep + \frac{\DPi_0} { \Ep} \right) G_0 (\bp, t ) \nonumber \\
& & \qquad +
\frac{\DPi_0} {2 \Ep}  \left( F_0 ( \bp, t ) + {\bar F}_0 ( \bp, t ) \right) \\
i \frac{\partial}{\partial t} {\bar G}_0 ( \bp, t) & = &                           
- \left( 2 \Ep + \frac{\DPi_0} { \Ep} \right) {\bar G}_0 (\bp, t )
\nonumber \\
& & \qquad  -
\frac{\DPi_0} {2 \Ep}  \left( F_0 ( \bp, t ) + {\bar F}_0 ( \bp, t ) \right) 
\end{eqnarray}

These coupled equations have a time-independent solution of the form
\begin{equation}
G_0 (\bp) =  {\bar G}_0 ( \bp ) = c \left( F_0 ( \bp ) + {\bar F}_0 ( \bp ) \right) 
\end{equation}
with 
\begin{equation}
c = - \frac{\DPi_0} { \Ep}  \left( 2 \Ep + \frac{\DPi_0} { \Ep} \right)^{-1}
\end{equation}
Thus if we define the particle mass $\mu$ so as to satisfy $\DPi_0 = 0$, then
two components of the Wigner functions $G_0 (\bp) $ and ${\bar G}_0 (\bp) $ vanish
for all $\bp$.
In this case,  $F_0 (\bp)$ may be interpreted as the momentum distribution of 
physical particle excitations.
 
To find time-dependent solutions, we write
\begin{eqnarray}
F_\pm (\bp. t ) & = & F_0 ( \bp, t) \pm {\bar F}_0 (\bp, t ) \\
G_\pm (\bp. t ) & = & G_0 ( \bp, t) \pm {\bar G}_0 (\bp, t ) 
\end{eqnarray}
and rewrite the equations as
\begin{eqnarray}
i \frac{\partial}{\partial t} F_+ ( \bp, t ) & = & - \frac{\DPi_0 (t)} { \Ep}  G_- ( \bp, t )  \\
i \frac{\partial}{\partial t} F_- ( \bp, t ) & = &  0 \\
i \frac{\partial}{\partial t} G_+ ( \bp, t ) & = &  
\left( 2 \Ep + \frac{\DPi_0 (t) } { \Ep} \right) G_- (\bp, t ) \\
i \frac{\partial}{\partial t} G_- ( \bp, t ) & = &   
 \left( 2 \Ep + \frac{\DPi_0 (t) } { \Ep} \right) G_+ (\bp, t ) 
+ \frac{\DPi_0} { \Ep}  F_+ ( \bp, t )  
\end{eqnarray}
These equations are to be solved together with the non-linear Klein-Gordon equation 
(\ref{nlKG0}) with
\begin{equation}
\DPi_0 (t) = \frac{\lambda}{2} \left[ \phi_0^2 (t) +  
\sum_\bp \frac{F_+ (\bp, t) + G_+ (\bp, t ) }{2\Ep} + m^2 - \mu^2 \right] 
\end{equation}
 
From these equations we can derive second order differential equation:
 \begin{eqnarray}
\frac{\partial^2}{\partial t^2} F_+ ( \bp, t ) & = &  
-  \left( \frac{\DPi_0} { \Ep} \right)^2 F_+ ( \bp, t )
-  \frac{\DPi_0} { E_{\bp}} \left( 2 \Ep + \frac{\DPi_0} { \Ep} \right) G_+ (\bp, t ) 
\nonumber \\
&& + i \frac{1} { \Ep} \frac{ d  \DPi_0 (t) }{d t }   G_- (\bp, t)  \\
\frac{\partial^2}{\partial t^2} G_+ ( \bp, t ) & = &  
- \frac{\DPi_0} { \Ep} \left( 2 \Ep + \frac{\DPi_0} { \Ep} \right) F_+ ( \bp, t )  
- \left( 2 \Ep + \frac{\DPi_0} { \Ep} \right)^2 G_+ (\bp, t ) \nonumber \\
&& - i \frac{1} { \Ep} \frac{ d  \DPi_0 (t) }{d t }    G_- (\bp, t)  
\end{eqnarray}
with
\begin{equation}
\frac{ d  \DPi_0 (t) }{d t }   =  \lambda \left[ \phi_0 (t) {\dot \phi}_0 (t) +
\sum_\bp G_- (\bp. t ) \right]
\end{equation}
We observe that each of these coupled equations looks like an equation of a forced 
oscillator.  For example the frequency of the oscillator $F_+$ is given by $\frac{\DPi_0} { \Ep}$
while the frequency of the oscillator $G_+$ is given by $2 \Ep + \frac{\DPi_0} { \Ep}$
so that $G_+$ oscillates much rapidly than $F_+$.    The coupling of these oscillators 
may produce an interesting interesting effects which may be studied by numerical 
integration.     

\section{Statistical equilibrium} 
In statistical equilibrium, we have the Bose distribution 
\begin{eqnarray}\label{bosedist}
F_0 (\bp) = f_{\rm eq.} ( \bp ) = \frac{1}{e^{\omega_\bp \beta} - 1}  
\end{eqnarray}
together with
\begin{equation}
G_0 (\bp) = g_{\rm eq.}  ( \bp ) =  0.
\end{equation}
where $\beta = 1/ k_B T$ is the inverse temperature.
The condition of vanishing $\DPi_0$ implies from (\ref{self})
\begin{equation}\label{mu20}
\frac{\lambda}{2} \left( \phi_0^2 + \la \tphi ^2 \ra_{\rm eq.} \right) + m^2 - \mu^2 = 0 .
\end{equation}
where the thermal fluctuation of the quantum field is given by the relation (\ref{phi2}) as
\begin{equation}
\la \tphi ^2 \ra_{\rm eq.} = \sum_\bp \frac{1}{\omega_\bp} f_{\rm eq.} ( \bp ) 
\end{equation}
On the other hand, the condensate amplitude $\phi_0$ in equilibrium should also satisfy 
the static non-linear Klein-Gordon equation,
\begin{equation}\label{phi0}
m^2  \phi_0  = - \frac{1}{3!} \lambda \left[ \phi_0^3 +  3 \la \tphi ^2 \ra_{\rm eq.} \phi_0 \right]
\end{equation}
From these two conditions, we find, for $m^2 < 0$, 
\begin{equation}\label{gap}
\mu^2 =  - 2 m^2 -  \lambda \sum_\bp \frac{1}{\omega_\bp} f_{\rm eq.} ( \bp ) 
\end{equation}
and 
\begin{equation}
\lambda \phi_0^2 = 3 \mu^2 
\end{equation}
We note that, since the thermal distribution $f_{\rm eq.}$ depends on the mass (gap) 
parameter $\mu$ through $\omega_\bp = \sqrt{ \bp^2 + \mu^2}$,  the equation (\ref{gap}) 
determines the mass parameter $\mu$ self-consistently as a function of the temperature.
This equation is called the {\it gap equation}~\cite{DJ74,BG77}. 
As the temperature increases the mass gap $\mu$ and the condensate amplitude $\phi_0$ 
decreases and vanishes at high temperature.

To find the behavior of the mass parameter as a function of the temperature
we can use the following formula~\cite{DJ74,RM98} for the integral of the bose
distribution function:
\begin{eqnarray}
\sum_\bp \frac{1}{\omega_\bp} f_{\rm eq.} ( \bp )  = 
\frac{1}{(2 \pi)^3} \int  \frac{d \bp}{\omega_\bp} \frac{1}{e^{\omega_\bp \beta} - 1}  
 =  \frac{\beta^{-2}}{2\pi^2}  I^{(2)}_- ( \mu \beta)  
\end{eqnarray}
where the dimensionless function $I^{(2)}_- ( x ) $ is given as
\begin{eqnarray}
I^{(2)}_- ( x )  & \equiv & \int_0^\infty \frac{k^2 d k}{\sqrt{k^2 + x^2}} \frac{1}{e^{\sqrt{k^2 + x^2}} - 1}
\nonumber \\
& = & \frac{\pi^2}{6} - \frac{\pi}{2}x - \frac{1}{4}x^2 \ln \frac{x}{4\pi} +
\left( \frac{1}{8} - \frac{1}{4} \gamma \right) x^2 - \frac{\zeta (3)}{32 \pi} x^4 + {\cal O}  ( x^6)
\nonumber \\
\end{eqnarray}
($\gamma = 0.57721 \cdots$ is Euler's number) .   The gap equation (\ref{gap}) has a solution
$\mu =0$ at $T_c$ determined by
\begin{equation}
(k_B T_c)^2 = - \frac{24}{\lambda} m^2 
\end{equation}
However the solution exhibits the behavior of the first order transition due to the non-analytic 
behavior of the function $I^{(2)}_- (x)$ \cite{RM98}.
This is a generic feature of the mean field approximation \cite{BG77} which may be a
theoretical artifact and may not survive the inclusion of correlations missing in the mean 
field approximation.  
An improvement of the mean field approximation has been proposed in \cite{CH98} 
and later it has been shown to lead to the second order transition in two-loop approximations 
\cite{Ch00}.     
 
\begin{figure}[htbp]
\begin{center}
\includegraphics{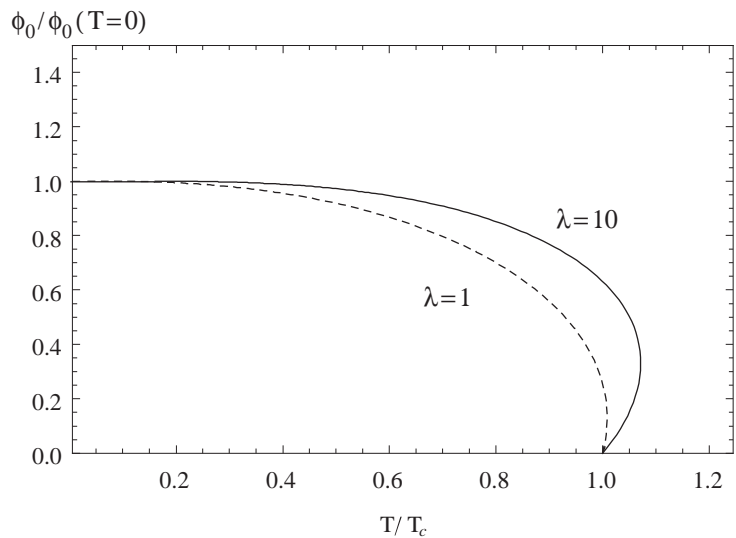}
\caption{Temperature dependence of the mass parameter $\mu$ for
$\lambda = 1$ (dashed line) and $\lambda = 10$ (solid line)}
\label{default}
\end{center}
\end{figure}

In Fig. 1 we plot $\phi_0$ as determined by (\ref{phi0}) as a function of temperature.  It shows
a behavior characteristic of the first order phase transition: there is a region $T_c < T < T_0$ 
where there are three solutions of (\ref{phi0}), one at $\phi_0 =0$ and other two at 
$\phi = \phi_1 \neq 0$ as well as $\phi = \phi_0 \neq 0$.  One expects that two solutions 
$\phi = \phi_0$ and $\phi = \phi_0$ correspond to two local minima of the effective potential 
$V ( \phi )$ while the other one $\phi = \phi_1$ corresponds to the local maximum of $V (\phi )$.  
Indeed we can show this explicitly by constructing the effective potential $V ( \phi) $ by
demanding that the local extremal condition $\partial V (\phi) /\partial \phi = 0 $ coincides 
with the condition (\ref{phi0}).    
\begin{equation}
V ( \phi ) = \int d \phi \left( m^2 \phi + \frac{\lambda}{3!} \phi^3 
+ \frac{\lambda}{2}\sum_\bp \frac{f_{\rm eq.} (\bp)}{\omega_\bp} \phi \right)
\end{equation}
The integration of the first two terms are trivial.  Beside irrelevant integration constant, 
which we may choose to be zero, we obtain
\begin{equation}
V ( \phi ) = \frac{1}{2} m^2 \phi^2 + \frac{\lambda}{4!} \phi^4 
+ \frac{\lambda}{2}  \int d \phi \phi \sum_\bp \frac{f_{\rm eq.} (\bp)}{\omega_\bp}
\end{equation}   
To carry out the remaining integral over $\phi$ we demand that the mass parameter $\mu$ 
depends on $\phi$ through the relation
\begin{equation}\label{mu2}
\mu^2 = m^2 + \frac{\lambda}{2} \left( \phi^2 + \la \phi ^2 \ra_{\rm eq.} \right) 
= m^2 + \frac{\lambda}{2} \left( \phi^2 + \sum_\bp \frac{f_{\rm eq.} (\bp)}{\omega_\bp} \right) 
\end{equation}
which appears to be equivalent to the condition (\ref{mu20}), but we assume that this
relation holds not only for the equilibrium value of $\phi$, namely at $\phi=\phi_0$, 
 but also for any value of $\phi$. 
Noting
\begin{equation}
\frac{\lambda}{2}  d \phi \phi = \left( 1 - \frac{\lambda}{2} \frac{\partial}{\partial \mu^2} \sum_\bp 
\frac{f_{\rm eq.} (\bp) }{\omega_\bp} \right) d \mu \mu   
\end{equation} 
which results from (\ref{mu2}), 
we transform the integration variable from $\phi$ to $\mu^2$:
\begin{eqnarray}
\frac{\lambda}{2}  \int d \phi \phi \sum_\bp \frac{f_{\rm eq.} (\bp) }{\omega_\bp} 
& = & \frac{1}{2}\int d \mu^2  \left( 1 - \frac{\lambda}{2} \frac{\partial}{\partial \mu^2} \sum_\bp 
\frac{f_{\rm eq.} (\bp) }{\omega_\bp} \right) \sum_{\bp'}  \frac{f_{\rm eq.} (\bp')}{\omega_{\bp'}}  
\nonumber \\
& = & \frac{1}{2}\int d \mu^2  \sum_{\bp}  \frac{f_{\rm eq.} (\bp) }{\omega_{\bp}} 
- \frac{\lambda}{8}  \left( \sum_{\bp}  \frac{f_{\rm eq.} (\bp) }{\omega_{\bp}} \right)^2 + \rm{const.}
\nonumber \\
\label{Integ1}
\end{eqnarray} 
Remaining integral over $\mu^2$ can be carried out as 
\begin{equation}\label{Integ2}
\frac{1}{2}\sum_\bp \int d \mu^2 \frac{f_{\rm eq.} (\bp) }{\omega_\bp} 
= \frac{1}{2\beta}\sum_\bp \ln \left[ 1 - \exp(- \beta \omega_\bp ) \right]   
= - \sum_\bp \frac{\bp^2}{3 \omega_\bp} f_{\rm eq.} (\bp) 
\end{equation} 
where  in deriving the final expression we performed integration by part in $p$ and 
omitted again the irrelevant integration constant.

To sum up, we find the following expression for the effective potential:
\begin{equation}\label{Veff}
V ( \phi ) = \frac{1}{2} m^2 \phi^2 + \frac{\lambda}{4!} \phi^4  
- \sum_\bp \frac{\bp^2}{3 \omega_\bp} f_{\rm eq.} (\bp)
- \frac{\lambda}{8}  \left( \sum_{\bp}  \frac{f_{\rm eq.} (\bp) }{\omega_{\bp}} \right)^2 + \rm{const.}
\end{equation}
where $\phi$ dependence of the last two terms are implicitly given from the $\phi$-dependence
of $\mu$. 
We note that this result coincides, in the neglect of the renormalization effects due to
the meson mass shift, with the effective potential obtained by Amelino-Camelia and 
Pi~\cite{ACP93} using 
the Cornwall-Jackiw-Tombolis (CJT) composite operator effective potential formalism~\cite{CJT74},  
We note that our mass parameter $\mu$ corresponds to the variational mass parameter $M$ 
of the CJT potential.  The third term may be interpreted as the pressure of an ideal gas of 
the quasi-particles obeying the dispersion $ \varepsilon = \omega_\bp = \sqrt{\bp^2 + \mu^2} $.   

The effective potential given by the formula (\ref{Veff}) takes in general a complex value 
since $\mu^2$ becomes negative at low temperature and at small value of $\phi$.   
At low temperatures the potential develops a second local minimum at non-zero value of 
$\phi$, which becomes the absolute minimum below certain temperature $T_1 ( T_c < T_1 < T_0 )$ where the first order transition takes place in equilibrium and the order parameter $\phi_0$ jumps 
discontinuously from $0$ to $\phi_0$ as the temperature is lowered.

\section{Slowly varying system: the Vlasov equation} 
In the presence of inhomogeneity the Wigner functions acquire non-vanishing elements
with $\bk \ne 0$.  If we assume that this inhomogeneity are due to the long-wavelength
fluctuations in the system, 
\begin{equation}\nonumber
k, q \ll p 
\end{equation} 
In this case we can obtain a familiar form of the Vlasov equation \cite{LP} 
from the equation of motion of the Wigner function by the procedure 
usually referred to as the gradient expansion \cite{BI02}.

To show this we make the following approximations: 
\begin{eqnarray}
\omega_{\bp + \bk/2}  + \omega_{\bp - \bk/2} && \simeq 2 \Ep \\
\omega_{\bp + \bk/2}  - \omega_{\bp - \bk/2} && \simeq 
\frac{\bp \cdot  \bk}{\omega_\bp}  
\end{eqnarray}
and
\begin{eqnarray}
\frac{1}{\sqrt{2\omega_{\bp+ \bk/2}}\sqrt{2\omega_{\bp + \bk/2+ \bq}} }
& \simeq & \frac{1}{2 \Ep} \left( 1 - \frac{\bp \cdot (\bk + \bq) }{\Ep^2} \right) \\
\frac{1}{\sqrt{2\omega_{\bp - \bk/2}}\sqrt{2\omega_{\bp - \bk/2 - \bq}} }
& \simeq & \frac{1}{2 \Ep} \left( 1 + \frac{\bp \cdot (\bk + \bq) }{\Ep^2} \right)
\end{eqnarray}
We also make use of the Taylor expansion of the Wigner functions:
\begin{eqnarray}
F ( \bp \pm \bq/2, \bk + \bq, t ) && \simeq F ( \bp , \bk + \bq, t ) \pm \frac{1}{2} \bq \cdot
\nabla_\bp F ( \bp , \bk + \bq, t ) \\
G ( \bp + \bq/2, \bk + \bq, t ) && \simeq G ( \bp , \bk + \bq, t ) + \frac{1}{2} \bq \cdot
\nabla_\bp G ( \bp , \bk + \bq, t ) \\
{\bar G} ( \bp - \bq/2, \bk + \bq, t ) && \simeq {\bar G}  ( \bp , \bk + \bq, t ) - \frac{1}{2} \bq \cdot
\nabla_\bp {\bar G}  ( \bp , \bk + \bq, t ) 
\end{eqnarray}

Then the equations of motion  (\ref{eomF}) of the Wigner function $F ( \bp, \bk, t )$ becomes
\begin{eqnarray}
\frac{\partial}{\partial t} F ( \bp, \bk, t ) & = &  
i \frac{\bp \cdot  \bk}{\omega_\bp}  F ( \bp, \bk, t )
+ i \sum_\bq \frac{ \DPi_\bq}{2\Ep }  \bq \cdot \nabla_\bp F ( \bp , \bk + \bq, t )
\nonumber \\
&& \qquad - i \sum_\bq \frac{ \DPi_\bq}{2 \Ep }  
\frac{\bp \cdot (\bk + \bq) }{\Ep^2} F ( \bp , \bk + \bq, t )
\nonumber \\
& & \quad  + i \sum_\bq \frac{ \DPi_\bq}{2 \Ep}  
\left( G ( \bp , \bk + \bq, t ) - {\bar G} ( \bp , \bk + \bq, t ) \right)
\nonumber \\
& &
\qquad  -  i \sum_\bq \frac{ \DPi_\bq}{2 \Ep }  
 \frac{\bp \cdot (\bk + \bq) }{\Ep^2}  
\left( G ( \bp , \bk + \bq, t )  + {\bar G} ( \bp , \bk + \bq, t ) \right) 
 \nonumber \\
 & &
\qquad  + i \sum_\bq \frac{ \DPi_\bq}{4 \Ep }  
 \bq \cdot \nabla_\bp 
\left( G ( \bp , \bk + \bq, t )  + {\bar G} ( \bp , \bk + \bq, t ) \right) 
 \nonumber \\
 \end{eqnarray}
On the right hand side, the first term and the second term are disguised forms of the
drift term and the Vlasov term respectively.   

To obtain more familiar form we make the Fourier transforms (42) and (43)  of 
the Wigner functions.  Then we find
\begin{eqnarray}
\frac{\partial}{\partial t} f ( \bp, \br, t )   & = & 
- \frac{\bp}{\Ep}  \cdot \nabla_\br f ( \bp, \br, t )
+ \nabla_\br \left(  \frac{ \DPi (\br, t ) }{2\Ep } \right) \cdot \nabla_\bp f ( \bp , \br, t )
\nonumber \\
& & \qquad + \frac{ \DPi (\br, t)}{2 \Ep } \frac{\bp}{\Ep^2}  \cdot \nabla_\br f ( \bp, \br, t )
\nonumber \\
& & \quad \qquad + i \frac{ \DPi (\br, t ) }{2 \Ep}  \left( g ( \bp , \br, t ) - {\bar g} ( \bp , \br, t ) \right)
\nonumber \\
& &
\qquad \quad + \frac{\DPi (\br, t)}{2 \Ep^2 }  
 \frac{\bp}{\Ep}  \cdot \nabla_\br 
\left( g ( \bp , \br, t )  + {\bar g} ( \bp , \br, t ) \right) 
 \nonumber \\
 & &
\qquad  \qquad + \nabla_\br \left( \frac{\DPi (\br, t ) }{4 \Ep }  \right)
\cdot \nabla_\bp 
\left( g ( \bp , \br, t )  + {\bar g} ( \bp , \br, t ) \right) 
 \nonumber \\
 \label{vlasov}
 \end{eqnarray}
Now it is clear that the first term is the drift term which describes the change of the particle 
position by the drift with the velocity $\bv_\bp = \bp/\Ep $.   The second term can be 
interpreted as the Vlasov term which represents the change of particle momentum due to
the continuous acceleration by the velocity dependent equivalent potential,
\begin{equation}\label{pot}
U_\bp ( \br, t ) = \frac{\DPi ( \br, t)}{2\Ep} = \frac{\Pi ( \br, t) + m^2 - \mu^2 }{2\Ep}
\end{equation}
acting on the particle.  The third term appears to be the 
correction to the drift term due to the local change of particle mass 
$\mu \to \mu' = \mu + \DPi (\br, t)$ which causes change in particle velocity.  
Other three terms are associated with the other components of the Wigner function and 
has no counter parts in non-relativistic Vlasov equation.

Noting that 
\begin{equation}
\frac{\partial \Ep}{\partial \bp} = \frac{\bp}{\Ep} , 
\end{equation}
and
\begin{equation}
\frac{\partial U_\bp (\br, t )}{\partial \bp} = - \frac{U_\bp (\br, t ) \bp}{\Ep^2} , 
\end{equation}
the above equation may be rewritten in a more compact form:
\begin{eqnarray}
\frac{\partial}{\partial t} f ( \bp, \br, t )   & &
+ \nabla_\bp \varepsilon ( \bp, \br, t ) \cdot \nabla_\br f ( \bp, \br, t )  
- \nabla_\br  \varepsilon ( \bp, \br, t )  \cdot \nabla_\bp f ( \bp , \br, t ) =
\nonumber \\
& &   \qquad  \qquad    i U_\bp (\br, t ) g_- ( \bp , \br, t ) 
 - \frac{1}{2} \nabla_\bp  U_\bp (\br, t)  \cdot \nabla_\br
g_+ ( \bp , \br, t )
 \nonumber \\
 &&
  \qquad  \qquad  \qquad  \qquad \qquad + \frac{1}{2} \nabla_\br  U_\bp (\br, t)  \cdot \nabla_\bp
g_+ ( \bp , \br, t ) 
 \nonumber \\
  \label{vlasov1}
 \end{eqnarray}
where
\begin{equation}
\varepsilon  ( \bp, \br, t ) = \Ep + U_\bp (\br, t ) 
\end{equation}
and
\begin{equation}
g_\pm ( \bp, \br, t ) = g ( \bp, \br, t ) \pm {\bar g} ( \bp, \br, t ).
\end{equation} 
The quantity $\varepsilon  ( \bp, \br, t ) $ plays the same role as the quasi-particle energy 
which appears in the kinetic equation of Landau's Fermi-liquid theory \cite{Lan57}.
\footnote{Relativistic extension of the general framework of the Landau Fermi-liquid theory 
has been made by Baym and Chin \cite{BC76} and applied by one (TM) of the present 
authors \cite{Mat81} to Walecka's relativistic mean field theory of cold dense nuclear 
matter \cite{Wal74}.}
We note that because of the relation (\ref{gstar-gbar}), $ g_ - ( \bp, \br, t )$ is a pure imaginary
function, while $g_+ ( \bp, \br, t)$, $f (\bp, \br, t)$, $U_\bp (\br, t )$, and
$\varepsilon  ( \bp, \br, t ) $  are all real functions.

In the long wavelength approximation, the equation of motion of $G ( \bp, \bk, t ) $
becomes
\begin{eqnarray}
\frac{\partial}{\partial t} G ( \bp, \bk, t ) & = &  
- 2 i \Ep  G ( \bp, \bk, t )
- i \sum_\bq \frac{ \DPi_\bq}{\Ep } G ( \bp , \bk + \bq, t )
\nonumber \\
& & \quad  - i \sum_\bq \frac{ \DPi_\bq}{2 \Ep}  
\left( F ( \bp , \bk + \bq, t ) + {\bar F} ( \bp , \bk + \bq, t ) \right)
\nonumber \\
& &
\qquad  +  i \sum_\bq \frac{ \DPi_\bq}{2 \Ep }  
 \frac{\bp \cdot (\bk + \bq) }{\Ep^2}  
\left( F ( \bp , \bk + \bq, t )  - {\bar F} ( \bp , \bk + \bq, t ) \right) 
 \nonumber \\
 & &
\qquad \quad  - i \sum_\bq \frac{ \DPi_\bq}{4 \Ep }  
 \bq \cdot \nabla_\bp 
\left( F ( \bp , \bk + \bq, t )  - {\bar F} ( \bp , \bk + \bq, t ) \right) 
 \nonumber \\
 \end{eqnarray}
which may be rewritten for the Fourier transforms in a compact form as
\begin{eqnarray}
\frac{\partial}{\partial t} g ( \bp, \br, t )  & + & 2 i \varepsilon (\bp, \br, t )  g ( \bp, \br, t ) 
 =   - i U_\bp (\br, t ) f_+ ( \bp , \br, t ) 
\nonumber \\
& &   \qquad  
 - \frac{1}{2} \nabla_\bp  U_\bp (\br, t)  \cdot \nabla_\br f_- ( \bp , \br, t )
 + \frac{1}{2} \nabla_\br  U_\bp (\br, t)  \cdot \nabla_\bp f_- ( \bp , \br, t ) 
 \nonumber \\
  \label{g-eq}
 \end{eqnarray} 
 where $f_{\pm}  (\bp, \br, t ) = f ( \bp, \br, t ) \pm {\bar f} ( \bp, \br, t )$. 
 Thus the kinetic equation for $g (\bp, \br, t )$ does not look like a Vlasov equation;
 it takes a form of a simple oscillator equation with the oscillator frequency 
 $2 \varepsilon (\bp, \br, t) $
 exposed to the external perturbation created by the particle distribution $f  (\bp, \br, t )$.
 Similar equation is derived for ${\bar g} ( \bp, \br, t) $ which is the complex conjugate of
 $g (\bp, \br, t )$.

Taking the complex conjugate of the Eq. (\ref{g-eq}), we obtain the kinetic equation for
${\bar g} (\bp, \br, t) = g^* (\bp, \br, t) $:
\begin{eqnarray}
\frac{\partial}{\partial t} {\bar g} ( \bp, \br, t )  & - & 2 i \varepsilon (\bp, \br, t )  {\bar g} ( \bp, \br, t ) 
 =   - i U_\bp (\br, t ) f_+ ( \bp , \br, t ) 
\nonumber \\
& &   \qquad  
 - \frac{1}{2} \nabla_\bp  U_\bp (\br, t)  \cdot \nabla_\br f_- ( \bp , \br, t )
 + \frac{1}{2} \nabla_\br  U_\bp (\br, t)  \cdot \nabla_\bp f_- ( \bp , \br, t ) 
 \nonumber \\
 \label{gbar-eq}
  \end{eqnarray} 
Adding or subtracting (\ref{g-eq}) and (\ref{gbar-eq}), we find for  the real function $g_+ (\bp,\br, t) $ 
and the pure imaginary function $g_- (\bp,\br, t) $
\begin{eqnarray}
\frac{\partial}{\partial t} g_+ ( \bp, \br, t )  & = & 
2 i \varepsilon (\bp, \br, t ) g_- ( \bp, \br, t ) 
- \nabla_\bp  U_\bp (\br, t)  \cdot \nabla_\br f_- ( \bp , \br, t )
\nonumber \\
& & \qquad  
 + \nabla_\br  U_\bp (\br, t)  \cdot \nabla_\bp f_- ( \bp , \br, t ) 
 \nonumber \\
 \label{g+} \\
 \frac{\partial}{\partial t} g_- ( \bp, \br, t )  & = & 
2 i \varepsilon (\bp, \br, t ) g_+ ( \bp, \br, t ) 
- 2 i U_\bp (\br, t ) f_+ ( \bp , \br, t ) 
 \label{g-}
  \end{eqnarray} 
  
In the long wavelength approximation, we also have   
\begin{eqnarray}\label{phi2'}
\la \hphi^2 (\br,t) \ra & \simeq & \sum_\bp
\frac{1}{2\omega_\bp} 
\left[ f ( \bp, \br, t) + {\bar f} ( \bp, \br, t) + g ( \bp, \br, t) + {\bar g} (\bp, \br, t) \right] 
\nonumber \\
& = & \sum_\bp
\frac{1}{2\omega_\bp} \left[ f_+ ( \bp, \br, t) +  g_+ ( \bp, \br, t) \right] 
\end{eqnarray}
which appears in the non-linear Klein-Gordon equation (\ref{nlKG})  for the condensate and in the
self energy term (\ref{Pi}) in the Vlasov equation.

\section{Dispersion relation of the excitations near equilibrium}

We now apply our coupled kinetic equations, consisting of the non-linear Klein-Gordon equation 
and the Vlasov equations, to find a dispersion relation of the excitations in the system 
near equilibrium. 
The dispersion relations of the excitations of the finite temperature system may be
computed using the propagator (two-time) formalism, either with real-times (Schwinger-Keldysh) 
formalism \cite{NS84} or with imaginary-times (Matsubara) formalism with subsequent analytic continuation \cite{Wel83}.   
Although such calculations have been performed by various authors\cite{CH98,PSS02,HMN03}numerical results have been presented focusing on the time-like region of the spectral functions.
We study with our kinetic theory the entire range of the $(\omega, k)$ plane including the 
space-like excitations of the system.
\footnote{We note that \cite{CH98} contains a general formula of the meson spectral function covering
all kinematic ranges including space-like region.  Their results are indeed very similar to ours, 
besides that the divergent 
vacuum loops are included with an elaborate temperature-dependent 
renormalization procedure and that the ``tree-level'' masses are used in the modified loop calculation, 
while we ignore the vacuum loops and the mass of our quasi-particle excitations is determined 
by the gap equation which includes the effect of one-loop diagrams in the language of 
conventional loop expansion.}   
As has been shown by one of the present authors \cite{Mat81} the calculation of the zero 
sound mode in the degenerate Fermi liquid using the relativistic Landau kinetic equation\cite{BC76}, 
similar to ours, reproduces the same result as the propagator theory in the long wavelength limits. 
We thus expect that our result may also reproduce the long wavelength behaviors of the 
excitations obtained from the propagator theory. 

For this purpose we assume that the distribution functions consist of uniform equilibrium term and
a small deviation from it: 
\begin{eqnarray}
f  ( \bp, \br, t )  & = & f_{\rm eq.} ( \bp ) + \delta f  ( \bp, \br, t )  \\
g  ( \bp, \br, t )  & = & g_{\rm eq.} ( \bp ) + \delta g  ( \bp, \br, t )  
\end{eqnarray}
where
\begin{eqnarray}
f_{\rm eq.} ( \bp ) = \frac{1}{e^{\omega_\bp \beta} - 1}  \qquad \mbox{\rm and} \qquad
g_{\rm eq.} ( \bp ) =  0.
\end{eqnarray}
We also assume the mean meson field consists of uniform equilibrium term and a deviation 
from it
\begin{equation}
\phi_c ( \br, t ) = \phi_0 + \delta \phi ( \br, t )
\end{equation}
where $\phi_0$ is determined by the static non-linear Klein-Gordon equation (\ref{phi0})
together with the solution of the gap equation (\ref{gap}) for the mass gap $\mu$.

Linearization of the non-linear Klein-Gordon equation with respect to $\delta \phi ( \br, t )$
yields 
\begin{eqnarray}\label{linKG}
\Box \delta \phi  (\br, t ) + m^2  \delta \phi  (\br, t ) 
& = &- \frac{1}{2} \lambda \left[ \phi_0^2 \delta \phi (\br, t ) + 
 \la \tphi ^2 \ra_{\rm eq.} \delta \phi (\br, t ) 
 +  \delta \la \tphi^2 ( \br, t ) \ra  \phi_0  \right] 
 \nonumber \\
\end{eqnarray} 
where the thermal fluctuation of the quantum field in equilibrium is given by 
\begin{equation}
\la \tphi ^2 \ra_{\rm eq.} = \sum_\bp \frac{1}{\omega_\bp} f_{\rm eq.} ( \bp ) 
\end{equation}
and the deviation $\delta \la \tphi^2 ( \br, t ) \ra$ is expressed in term of the
small deviation of the distribution function from equilibrium value.
\begin{eqnarray}\label{deltaphi2}
\delta \la \hphi^2 (\br,t) \ra & = & \sum_\bp
\frac{1}{2\omega_\bp} 
\left[ \delta f ( \bp, \br, t) + \delta {\bar f} ( \bp, \br, t) 
+ \delta g ( \bp, \br, t ) + \delta {\bar g} (\bp, \br, t) \right] 
\nonumber \\
 & = & 
 \sum_\bp
\frac{1}{2\omega_\bp} 
\left[ \delta f ( \bp, \br, t) + \delta f ( - \bp, \br, t) 
+ \delta g ( \bp, \br, t ) + \delta g^* (\bp, \br, t) \right] 
\nonumber \\
\end{eqnarray}
where in deriving the last line we have used 
\begin{eqnarray}
\delta {\bar f} ( \bp, \br, t) & = & \delta f ( -\bp, \br, t )  \\
\delta {\bar g} ( \bp, \br, t) & = & \delta g^* ( \bp, \br, t ) 
\end{eqnarray}
which follow from (\ref{fbar}) and (\ref{gbar}) respectively.

By the linearization with respect to the small increments,
$\delta f  ( \bp, \br, t )$,  $\delta g ( \bp, \br, t )$, the Vlasov equations become
\begin{eqnarray}
\frac{\partial}{\partial t} \delta f ( \bp, \br, t )   & + &
\nabla_\bp \omega_\bp \cdot \nabla_\br \delta f ( \bp, \br, t )  
- \nabla_\br \delta U_\bp ( \br, t ) \cdot \nabla_\bp f_{\rm eq.} ( \bp ) = 0
\nonumber \\
\label{lin-vlasov1} \\
\frac{\partial}{\partial t} \delta g ( \bp, \br, t )  
& + & 2 i \omega_\bp \delta  g ( \bp, \br, t ) 
 =  - i 2 \delta U_\bp (\br, t ) f_{\rm eq.} ( \bp ) 
\nonumber \\
\label{lin-vlasov2}
\end{eqnarray}
The right hand side of the Vlasov equation (\ref{vlasov1}) vanishes because 
$\DPi (\bp, \br, t )$ vanishes in equilibrium. 
The fluctuation in the mean field potential $\delta U_\bp ( \br, t )$ which appears in these
equations is related to the fluctuations of the condensate amplitude and the distribution functions    
\begin{eqnarray}\label{deltaU}
\delta U_\bp ( \br, t ) =  \frac{1}{2\omega_\bp}  \delta \Pi (\br, t) 
= \frac{\lambda}{2\omega_\bp}  \left( 2 \phi_0 \delta \phi (\br, t ) 
+ \delta \la \tphi^2 (\br, t ) \ra \right) 
\end{eqnarray}
The linearized Klein-Gordon equation (\ref{linKG}) and the linearized Vlasov equations
(\ref{lin-vlasov1}) and (\ref{lin-vlasov2}), supplimented by the ``constitutive relations" (\ref{deltaU}) and (\ref{deltaphi2}), form a closed set of equations to determine
a small fluctuation propagating the system in equilibrium. 

Noting that $\delta \phi (\br, t )$ and $\delta f ( \bp, \br, t )$ are real functions,  
we seek a solution in the form:
\begin{eqnarray}
\delta \phi (\br, t ) & = &  \delta \phi e^{i ( \bk \cdot \br - \omega_+ t )} 
+ \delta \phi^* e^{ - i ( \bk \cdot \br - \omega_- t )} , \\
\delta f (\bp, \br, t ) & = & \delta f_\bp e^{i ( \bk \cdot \br - \omega_+ t )}
+ \delta f_\bp^* e^{ - i ( \bk \cdot \br - \omega_- t )} , \\
\delta g (\bp, \br, t ) & = & \delta g_\bp e^{i ( \bk \cdot \br - \omega_+ t )} ,\\
\delta {\bar g} (\bp, \br, t ) & = & \delta g^* (\bp, \br, t ) = \delta g^*_\bp e^{- i ( \bk \cdot \br - \omega_- t )} ,
\end{eqnarray}
where we have introduced the Landau prescription~\cite{LP}, 
\begin{equation}
\omega_\pm = \omega \pm i \epsilon
\end{equation}
with a positive infinitesimally small constant $\epsilon$, to set an adiabatic 
switching-on of the fluctuation, namely 
$\delta f (\bp, \br ,t) $, $\delta g (\bp, \br ,t)$ all 
vanish slowly as $t \to - \infty$.
The constitutive relations for the fluctuations (\ref{deltaphi2}) and the mean 
field potential (\ref{deltaU}) now read
\begin{eqnarray}
\delta \la \hphi^2 (\br,t) \ra  & = & 
 \sum_\bp
\frac{1}{2\omega_\bp} 
\left( \delta f_\bp + \delta f_{ - \bp} + \delta g_\bp + \delta g^*_{-\bp} \right) 
e^{ i (\bk \cdot \br - \omega_+ t )} 
\nonumber \\
& & \qquad + \sum_\bp
\frac{1}{2\omega_\bp} 
\left( \delta f^*_\bp + \delta f^*_{ - \bp} + \delta g^*_\bp +  \delta g_{-\bp} \right ) 
e^{- i (\bk \cdot \br - \omega_- t )} 
\nonumber \\
\end{eqnarray}
and 
\begin{eqnarray}
\delta U_\bp ( \br, t ) 
& = & \frac{\lambda}{2\omega_\bp}  \left[ 2 \phi_0 \delta \phi 
+ \sum_\bp \frac{1}{2\omega_\bp} 
\left( \delta f_\bp + \delta f_{ - \bp} + \delta g_\bp + \delta g^*_{-\bp} \right) \right]
e^{ i (\bk \cdot \br - \omega_+ t )} 
\nonumber \\
& & \quad + \frac{\lambda}{2\omega_\bp}  \left[ 2 \phi^*_0 \delta \phi^*
+ \sum_\bp \frac{1}{2\omega_\bp} 
\left( \delta f^*_\bp + \delta f^*_{ - \bp} + \delta g^*_\bp + \delta g_{-\bp} \right) \right]
e^{ - i (\bk \cdot \br - \omega_- t )} 
\nonumber \\
\end{eqnarray}
respectively.

Using these relations, we find from the linearized Klein-Gordon equation (\ref{linKG}),
\begin{equation}\label{deltaphi}
\left[  - \omega_+^2 + \bk^2 +  \mu^2 \right] 
\delta \phi  
=  - \frac{1}{2} \lambda \phi_0 \sum_\bp \frac{1}{2\omega_\bp} 
\left[ \delta f_\bp + \delta f_{-\bp} + \delta g_\bp + \delta g^*_{-\bp} \right] ,
\end{equation}
and its complex conjugate relation, where we have used the equilibrium relation
\begin{equation}
\mu^2 = m^2
+ \frac{ \lambda}{2} \left( \phi_0^2  +  \la \tphi^2 \ra_{\rm eq.} \right)
\end{equation}
Also, from the linearized Vlasov equations we obtain
\begin{eqnarray}
( \omega_+ - \bv_\bp \cdot \bk ) \delta f_\bp
& = &   \frac{\lambda}{2\omega_\bp}  (  \bv_\bp \cdot \bk ) \beta  
\left[ 2 \phi_0 \delta \phi + \sum_{\bp'} \frac{1}{2\omega_{\bp'} }  
\left( \delta f_{\bp'} + \delta f_{ - \bp'} + \delta g_{\bp'} + \delta g^*_{-\bp'} \right) \right] 
\nonumber \\
& & \qquad \qquad \times
\left( 1 + f_{\rm eq.} ( \bp ) \right) f_{\rm eq.} ( \bp ) 
\nonumber \\
\label{deltaf}
\end{eqnarray}
and 
\begin{eqnarray}
\left( - \omega_+ + 2 \omega_\bp \right) \delta g_\bp
& = &  - \frac{\lambda}{\omega_\bp}  \left[ 2 \phi_0 \delta \phi 
+ \sum_{\bp'} \frac{1}{2\omega_{\bp'}} 
\left( \delta f_{\bp'} + \delta f_{ - \bp'} + \delta g_{\bp'} + \delta g^*_{-\bp'} \right) \right] f_{\rm eq.} ( \bp ) 
\nonumber \\
\label{deltag}
\end{eqnarray}
where we have used
\begin{eqnarray}
\nabla_\bp \omega_\bp & = & \frac{\bp}{\omega_\bp} = \bv_\bp \\
\nabla_\bp f_{\rm eq.} (\bp) & = & - \bv_\bp \beta \left( 1 +  f_{\rm eq.} (\bp) \right)
 f_{\rm eq.} (\bp) 
\end{eqnarray}
Using (\ref{deltaphi}), we eliminate $\delta \phi$ in (\ref{deltaf}) and (\ref{deltag}) and find
\begin{eqnarray}
( \omega_+ - \bv_\bp \cdot \bk )\delta f_\bp & = & 
 \frac{\lambda}{2\omega_\bp}  (  \bv_\bp \cdot \bk ) \beta 
\left( 1 + f_{\rm eq.} ( \bp ) \right) f_{\rm eq.} ( \bp ) 
\nonumber \\
& & \qquad \times
\left( 1 + \frac{\lambda \phi_0^2}{\omega_+^2 - \bk^2 - \mu^2}  \right)  {\cal F}
\label{df}
\\
\left( \omega_+ - 2 \omega_\bp \right) \delta g_\bp
& = &   \frac{\lambda}{\omega_\bp} f_{\rm eq.} ( \bp )  
\left( 1 + \frac{\lambda \phi_0^2}{\omega_+^2 - \bk^2 - \mu^2}  \right)
{\cal F} 
\label{dg}
\end{eqnarray}
where we have introduced the notation
\begin{equation}\label{F}
{\cal F} = \sum_{\bp} \frac{1}{2\omega_{\bp}} 
\left( \delta f_{\bp} + \delta f_{ - \bp} + \delta g_{\bp} + \delta g^*_{-\bp} \right)  
\end{equation}
for the deviation of the fluctuation. 
Solving (\ref{df}) and (\ref{dg}) for $\delta f_\bp$ and $\delta g_\bp$ respectively,
and inserting the results into (\ref{F}) we obtain the relation
\begin{eqnarray}\label{OmegaF}
{\cal F} = \Omega (\omega_+, \bk ) {\cal F}
\end{eqnarray}
where 
\begin{eqnarray}\label{Phi}
\Omega (\omega_+, \bk ) & = &  \frac{\lambda}{2}
\left( 1 +  \frac{\lambda \phi_0^2}{\omega_+^2 - \bk^2 - \mu^2}  \right) 
 \int  \frac{d^3 \bp}{(2 \pi)^3}  \frac{1}{\omega^2_\bp} 
\nonumber \\
& & 
\qquad 
\times \left[  \frac{ 4 \omega_\bp f_{\rm eq.} ( \bp ) }{ \omega_+^2 - 4 \omega_\bp^2 } 
+ \frac{  \beta ( \bv_\bp \cdot \bk)^2 } {\omega_+^2 - ( \bv_\bp \cdot \bk )^2} 
\left( 1 + f_{\rm eq.} ( \bp ) \right)  f_{\rm eq.} ( \bp ) \right]  ~.
 \nonumber \\
\end{eqnarray}

Applying the prescription,
\begin{equation}
\lim_{\epsilon \to +0} \frac{1}{\omega - \omega_0 + i \epsilon} = 
{\cal P} \frac{1}{\omega - \omega_0} - i \pi \delta (\omega - \omega_0 )
\end{equation}
with $\cal P$ implying that the principal part should be taken in integration, 
we find that $\Omega (\omega_+, \bk )$ consists of the
real and imaginary parts:
\begin{equation}
\Omega (\omega_+, \bk )  =  \Omega_1 (\omega, \bk ) + i \Omega_2 (\omega, \bk )
\end{equation}
The real part is given by  
\begin{eqnarray}
\Omega_1 (\omega, \bk ) 
& = & - \frac{\lambda}{2}
\left( 1 +  {\cal P} \frac{\lambda \phi_0^2}{\omega^2 - \bk^2 - \mu^2}  \right) \Phi_1 (\omega, k )
\end{eqnarray}
with
\begin{eqnarray}\label{Phi1}
\Phi_1 (\omega, k ) & = & -
{\cal P} \int  \frac{d^3 \bp}{(2 \pi)^3}  \frac{1}{\omega^2_\bp} 
\left[  \frac{ 4 \omega_\bp f_{\rm eq.} ( \bp ) }{ \omega^2 - 4 \omega_\bp^2 } 
+ \frac{  \beta ( \bv_\bp \cdot \bk)^2 } {\omega^2 - ( \bv_\bp \cdot \bk )^2} 
\left( 1 + f_{\rm eq.} ( \bp ) \right)  f_{\rm eq.} ( \bp )
  \right]  ~,
 \nonumber 
\\
\end{eqnarray}
while the imaginary part is given by
\begin{eqnarray}
\Omega_2 ( \omega, \bk ) & = & - \frac{\lambda}{4}
\left( 1 + {\cal P} \frac{\lambda \phi_0^2}{\omega^2 - \bk^2 - \mu^2}  \right)  \Phi_2 (\omega, k ) 
\nonumber \\
&  & \qquad \quad  +  \frac{\lambda^2 \phi_0^2}{4\omega_\bk}
\pi \left( \delta ( \omega - \omega_\bk) 
- \delta ( \omega + \omega_\bk ) \right) \Phi_1 (\omega_\bk , k ) 
\nonumber \\
& & 
\end{eqnarray}
with
\begin{eqnarray}
\Phi_2 (\omega, k ) & = & 
 \int  \frac{d^3 \bp}{(2 \pi)^3}  \frac{1}{\omega^2_\bp} \left[  
 2 \pi \left( \delta ( \omega - 2 \omega_\bp )
- \delta ( \omega + 2 \omega_\bp ) \right)  f_{\rm eq.} ( \bp ) \right.
\nonumber \\
& &  \left. \quad 
+ \pi \beta  \bv_\bp \cdot \bk \left( \delta ( \omega - \bv_\bp \cdot \bk )
- \delta ( \omega + \bv_\bp \cdot \bk ) \right)  
\left( 1 + f_{\rm eq.} ( \bp ) \right)  f_{\rm eq.} ( \bp )
  \right]
 \nonumber  \\
 \end{eqnarray}
where the integration can be carried out analytically, yielding
\begin{eqnarray}\label{Phi2}
\Phi_2 ( \omega, \bk ) 
& = &
  \frac{1}{8\pi^2} 
\frac{\omega}{k} \frac{1}{e^{\frac{\mu\beta}{\sqrt{1 - (\omega/k)^2}}} -1}  \theta ( k - \omega) 
 + \frac{\sqrt{\omega^2 - 4 \mu^2}}{2 \pi \omega } \frac{1}{e^{\omega \beta/2} - 1}
\theta (\omega - 2\mu)
\nonumber \\
\end{eqnarray}
for $\omega > 0$.  The values of this function for $\omega < 0$ can be found by noting that it 
is an odd function of $\omega$. 

The first term in (\ref{Phi2}) corresponds to the space-like (scattering) continuum with 
$\omega = \omega_{\bp + \bk} - \omega_\bp \simeq \bv_\bp \cdot \bk < k$. 
while the second term corresponds to the continuum of thermally induced pair 
creation/annihilation of mesons with energy $\omega > 2 \omega_{\bk/2} \simeq 2 \mu$, 
The function $\Phi_2 (\omega, k)$ (and $\Omega_2 (\omega, k)$) has non-vanishing supports 
in the regions on $(\omega, k)$ plane as indicated by the shaded areas in Fig. 2.  
We plotted in Fig. 3 $\Phi_1 (\omega, k)$ and $\Phi_2 (\omega, k)$ as functions of $\omega/k$ 
at $k = 0.5 \mu$ for two different values of $\mu/T$.

\begin{figure}[htbp]
\begin{center}
\includegraphics{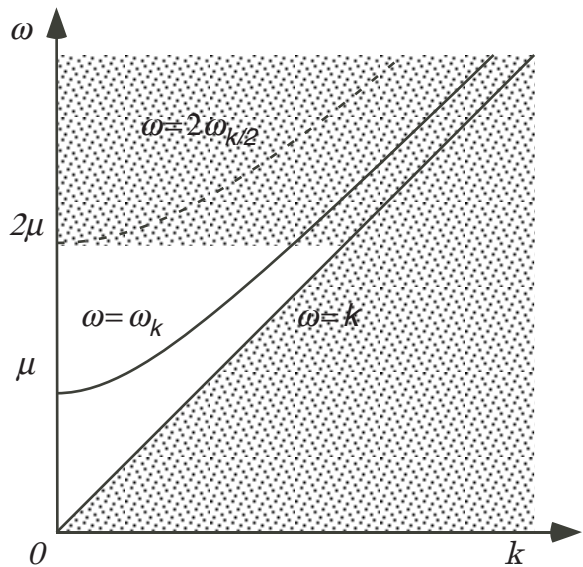}
\caption{The regions where $\Omega_2 (\omega, k)$ and $\Phi_2 (\omega, k)$ have 
non-vanishing value is shown by shaded areas (two-quasi-particle continua). 
The solid hyperbola corresponding to the meson poles ($\omega =  \omega_{\bk}$) 
on which $\Omega_2 (\omega, k)$ has a $\delta$-function singularity in the low temperature
phase.  The shaded area in the time-like region below the curve $\omega = 2 \omega_{\bk/2}$
is only an artifact of the long wavelength approximation. }
\label{default}
\end{center}
\end{figure}

\begin{figure}[htbp]
\begin{center}
\includegraphics{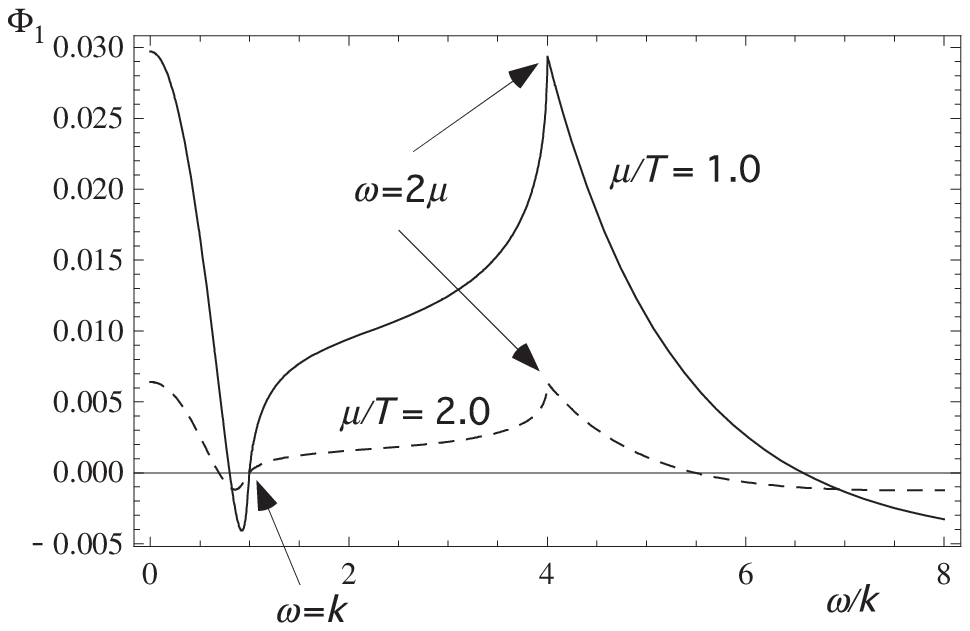}
\includegraphics{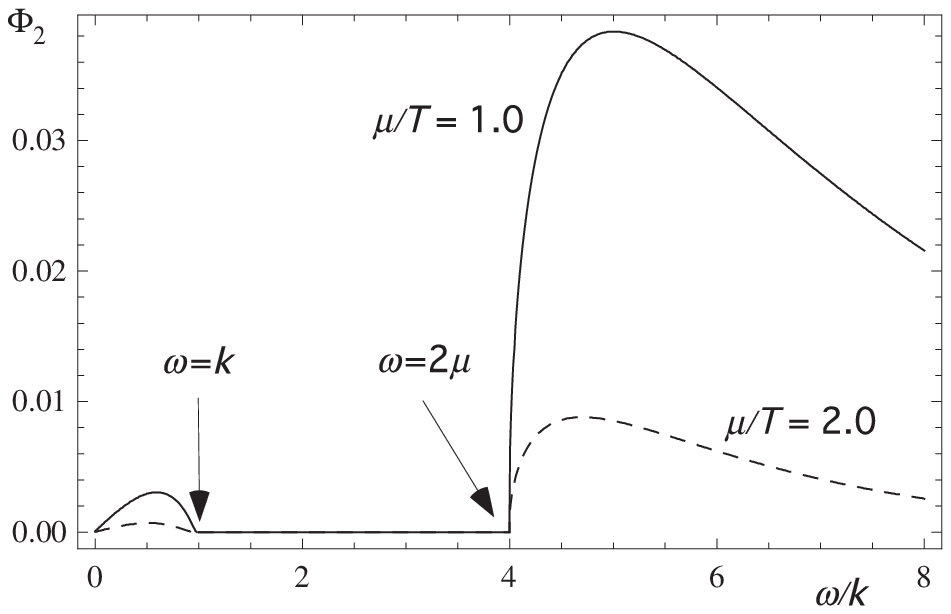}
\caption{$\Phi_1 ( \omega, k)$ and $\Phi_2 ( \omega, k)$ as a function of $\omega/k$  at  
$k/\mu = 0.5$ and $\mu/T = 1$ (dashed line) , $\mu/T =2$ (solid line).  Both $\Phi_1$ 
and $\Phi_2$ vanish at $\omega = k$.  $\Phi_2$ also vanishes at the pair creation 
threshold $\omega = 2 \mu$ (in the long wavelength approximation) while $\Phi_1$ has 
a cusp at this point.}
\label{default}
\end{center}
\end{figure}

The condition
\begin{equation}\label{dispersion}
1 = \Omega ( \omega, \bk ) 
\end{equation}
which follows from (\ref{OmegaF}) determines the dispersion relation of
a possible long wavelength collective excitation of the system. 

Before examining the solutions of the dispersion relation (\ref{dispersion}).
we note that at zero temperature, namely in the absence of quasi-particle
excitations, the only solution of our coupled kinetic equations is the one 
which satisfies the linearized Klein-Gordon equation (\ref{linKG}) or
(\ref{deltaphi}) with $\delta f = \delta g = 0$ and 
$\mu = \sqrt{ m^2 + \lambda \phi_0^2/2}$. 
This gives a simple meson pole $\omega = \omega_{\bk} = \sqrt{\bk^2+\mu^2}$ 
which appears in the time-like region ($\omega > k$).   
There is no space-like mode of excitations in the absence of quasi-particle excitations
due to the mass gap ($\Delta \omega = 2 \mu$) for the excitations of the vacuum.

The situation is different in the case of the ordinary Bose-Einstein Condensate (BEC)
which possesses, even at zero temperature, a low-lying acoustic mode 
(Bogoliubov phonon) in the mean-field approximation to repulsive two-body 
contact interaction\cite{Bog47,PS02}.   
The ordinary BEC contains space-like quasi-particle excitations even at absolute 
zero temperature which corresponds to the excitation of one of the zero energy 
particles forming the condensate.  The Bogoliubov phonon is a collective excitation of 
such space-like quasi-particle excitation modes modified by the repulsive 
mean-field interaction.   

At finite non-zero temperature, the system contains continuum of the space-like 
excitations for all $(\omega, \bk)$ satisfying $\omega < k$ as signified by the 
non-vanishing value of $\Omega_2 (\omega, \bk )$ in addition to the continuum
in the time-like region $\omega > 2 \mu$.  (See Fig. 2) Any solution of (\ref{dispersion}) in 
the space-like region therefore is subject to the collisionless dissipation
known as the Landau damping\cite{LP}. 

We plot the function $\Omega_1  (\omega, \bk )$ in Fig. 4 as a function of $\omega$ 
at fixed value of $k = 0.5 \mu$.  
In the low temperature phase, it contains a singularity at the position of the mesonic 
pole at $\omega = \sqrt{k^2 + \mu^2}$ while this singularity disappears in the 
high temperature phase.  
In the presence of the condensate, a disturbance in the quasi-particle distribution 
may be absorbed into an excitation of the condensate which then propagates with 
the mesonic dispersion relation and is converted back again to the quasi-particle
excitations.  This coupling between the condensate and the quasiparticle excitations 
generates a collective mode in the low temperature phase. 
\footnote{This mode may be compared to the collisionless acoustic mode (the quasi-particle 
sound \cite{NP90} ) which appears in the superfluid $^4$He due to the excitation 
of the condensate coupled with quasi-particle excitations.} 
In the high temperature phase, the quasi-particle excitations couple each other only
through their direct interaction, hence no mixing with single mesonic pole. 

The high temperature behavior of $\Omega_1$ is the same as that of the function 
$\Phi_1$: it is negative at $\omega =0$ and increases with $\omega$ and reaches 
a positive maximum slightly below $\omega = k$.
In the low temperature phase,  $\Omega_1  (\omega, \bk )$ reverts the sign for 
$\omega$ below this singularity.   
One may interpret that effective coupling strength  
\begin{equation}\label{lambda'}
\lambda' (\omega, k) = \lambda \left( 1 +  \frac{\lambda \phi_0^2}{\omega^2 - \bk^2 - \mu^2}  \right) 
= \lambda \left( 1 +  \frac{3 \mu^2}{\omega^2 - \bk^2 - \mu^2}  \right) 
\end{equation}
changes its sign below the meson pole.  
\footnote{This behavior reminds us of the change of the effective two-body interaction 
in the Bose-Einstein condensates as a function of the external magnetic field 
due to the coupling to the intermediate atomic resonance state, the phenomenon
known as the Feshbach resonance \cite{PS02}.}
Note that here we have used the relation 
$\lambda \phi_0^2 = 3 \mu^2$ (111) for the condensate amplitude. 
We note that  $\Omega_1  (\omega, k )$ always vanishes
at $\omega = k$ since $\Phi_1  (k, k ) = 0$. 
The cusp at $\omega = 2 \mu$ appears at the threshold of the two quasiparticle
creation.  The condition (\ref{dispersion}) is fulfilled only near the meson pole 
in the low temperature phase and it gives a shift of the mesonic excitation spectrum.
We could not find any additional solution satisfying (\ref{dispersion}).   

We like to note here that without the coupling to the fluctuations to pair creation
or annihilation, we would not have gotten the first term in the function $\Phi_1 (\omega, k)$
and then $\Omega_1 (\omega, k)$ would have become 1 at $\omega/k > 1$ creating a
undamped tachyonic sound mode.  Hence the off-diaginal Wigner functions plays an important 
role in making our framework consistent with causality. 

\begin{figure}[htbp]
\begin{center}
\includegraphics{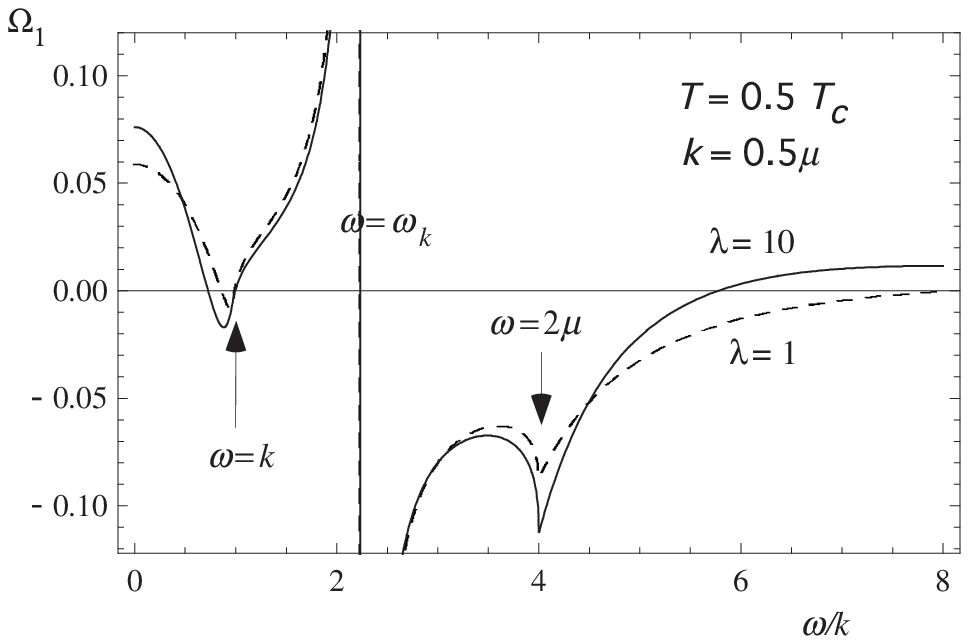}
\includegraphics{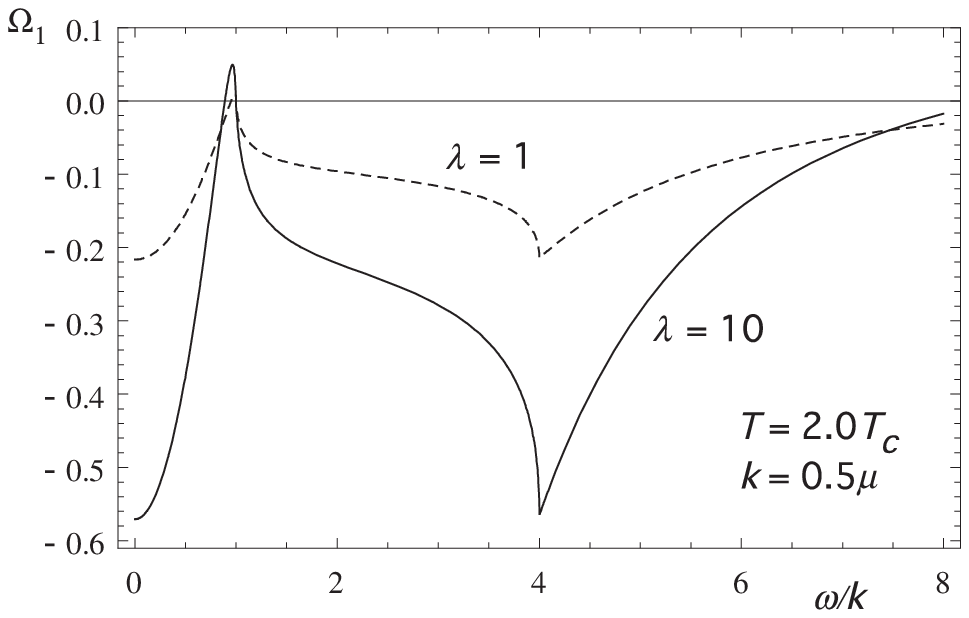}
\caption{$\Omega_1 ( \omega, k)$ as a function of $\omega/k$  at fixed $k$}
\label{default}
\end{center}
\end{figure}

We introduce the "response function" defined by
\begin{equation}\label{response}
R (\omega, k ) =  - {\rm Im} \left[ \frac{1}{1 - \Omega (\omega, \mu ) }  \right]
=  - \frac{ \Omega_2 (\omega, \mu ) }{\left[1 - \Omega_1(\omega, \mu ) \right]^2 
+ \left[ \Omega_2 (\omega, \mu ) \right]^2} 
\end{equation}
It is plotted in Fig. 5 as a function of $\omega$ and compared with the "bare" 
response function given by 
\begin{equation}\label{response}
R_0 (\omega, k ) =  - \Omega_2 (\omega, \mu )  
\end{equation}
The response function in the space-like momentum region gives the dynamic form 
factor of the cross section of the scattering of particles coupled to the excitations 
of the system \cite{VH54,PN66}, while its time-like component may give the rate of
pair annihilation of the quasi-particles.  
We note that in the low temperature phase the response function changes its sign
below the meson pole due to the sign change of the effective coupling 
(\ref{lambda'}).  
We observe some enhancement (depletion) of the strength in the time-like pair 
annihilation near threshold and small depletion (enhancement ) in the space-like
region in the low (high) temperature phase. 

\begin{figure}[htbp]
\begin{center}
\includegraphics{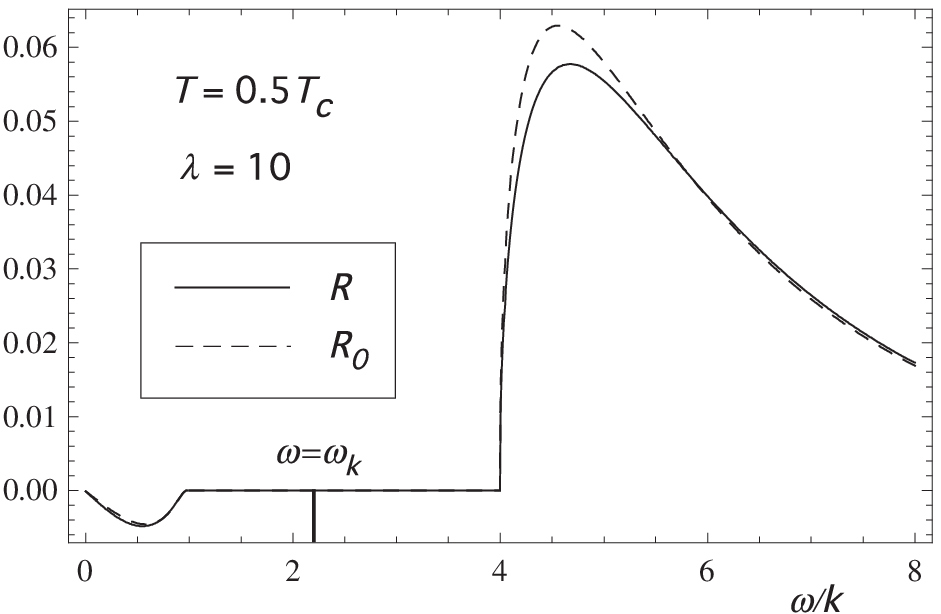}
\includegraphics{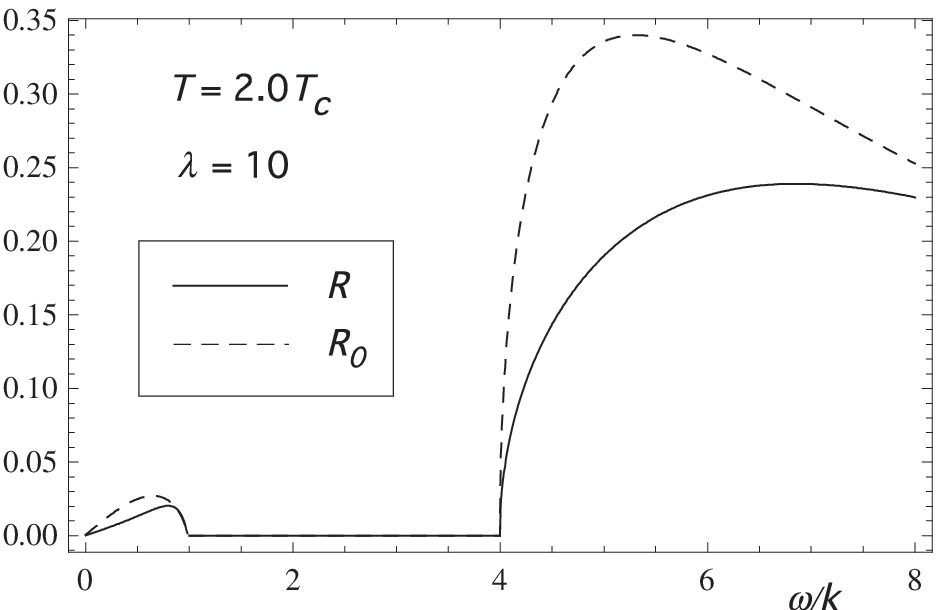} 
\caption{Plot of the response function $R (\omega, k)$ (\ref{response}) (solid lines) at $k/\mu = 0.5$.
For comparison the bare response function $R_0 (\omega, k) = - \Omega_2 (\omega, k)$ 
is also shown (dashed lines). Note that in the low temperature phase $R$ changes sign 
in the space-like region and at the meson pole ($\omega \simeq \omega_\bk$) due to the 
sign change of the effective coupling strength (\ref{lambda'}).}
\label{default}
\end{center}
\end{figure}

As we have noted, in the low temperature phase with non-vanishing condensate 
amplitude $\phi_0$, the mesonic excitations couple with the quasi-particle excitations 
and this give the additional shift of the meson mass from the one obtained by the
gap equation.   
The shift of the meson pole from the bare spectrum 
$\omega = \omega_\bk = \sqrt{k^2 + \mu^2}$ may be computed by the condition 
\begin{equation}
\omega^2 - k^2 - \mu^2 - ( \omega^2 - k^2 - \mu^2) \Omega (\omega, k) = 0
\end{equation}
which in the long wavelength limit ( $k = 0$ ) gives a solution at $\omega^2 = \mu'^2$.
The shift of the meson mass $\Delta \mu^2  = \mu'^2 - \mu^2$ is determined by
\begin{eqnarray}\label{mass-shift}
\Delta \mu^2   =   \mu'^2 - \mu^2 & =  & 2 \lambda
\left(  \Delta \mu^2  + 3 \mu^2  \right) 
\int  \frac{d^3 \bp}{(2 \pi)^3\omega_\bp} 
\frac{ f_{\rm eq.} ( \bp ) }{  \mu^2 + \Delta \mu^2 - 4 \omega_\bp^2 }  
\end{eqnarray}
where we have used $\lambda \phi_0^2 = 3 \mu^2$. 
In the case of $| \Delta \mu^2 |  << \mu^2$ this may be solved approximately
\begin{equation}\label{mass-shift0}
\Delta \mu^2  \simeq   6 \lambda  \mu^2   
\int  \frac{d^3 \bp}{(2 \pi)^3\omega_\bp}  
\frac{  f_{\rm eq.} ( \bp ) }{  \mu^2  - 4 \omega_\bp^2 }  
\left/  \left(  1 +  2 \lambda  \int  \frac{d^3 \bp'}{(2 \pi)^3\omega_{\bp'}}  
\frac{ f_{\rm eq.} ( \bp' ) }{  \mu^2  - 4 \omega_{\bp'}^2 }  \right) \right.
\end{equation}

The equation (\ref{mass-shift}) may be solved by iteration starting from
the first approximation (\ref{mass-shift0}) by inserting it for $\Delta \mu^2$ in 
the integrand on the right hand side.    
We plot in Fig. 6 the meson mass shift determined by this method. 
The meson mass becomes smaller as $T$ increases and eventually becomes
zero at the temperature which satisfies the condition 
\begin{eqnarray}\label{spinodal}
1 =   \lambda \int  \frac{d^3 \bp}{(2 \pi)^3 } \frac{1}{\omega_\bp^3}  f_{\rm eq.} ( \bp )   
\end{eqnarray}
The vanishing of the effective mass of mesonic excitation may be identified 
as the onset of instability in the metastable low temperature phase by small fluctuation 
(spinodal decomposition).  
The same instability occurs when one approaches to $T = T_c$ from high temperature phase.

We note that the temperature $T_{\rm sp}$ at the spinodal point given 
by (\ref{spinodal}) does not coincide with the "backbending" temperature $T_1$ 
where $d \mu/ dT$ diverges.  The latter temperature 
is determined by the condition
\begin{equation}
1 =   \lambda \int  \frac{d^3 \bp}{(2 \pi)^3 } \frac{1}{2  p^2 \omega_\bp}  f_{\rm eq.} ( \bp ) 
\end{equation}
which can be obtained from the gap equation (\ref{gap}). 
We found that $T_1$ is slightly above $T_{\rm sp}$ and the unstable region appears
associated with the lower solutions $\mu$ of the gap equation. 
These unstable solutions thus appear only in the region which is not easily accessible
and may well be considered as another pathology of the mean field approximation.
\footnote{These two temperatures may coincide with $T_c$ in the case of the second order 
transition of the Landau-type.  Such behavior has been obtained in \cite{Ch00} by a
temperature-dependent loop expansion method formulated in \cite{CH98}. }

\begin{figure}[htbp]
\begin{center}
\includegraphics{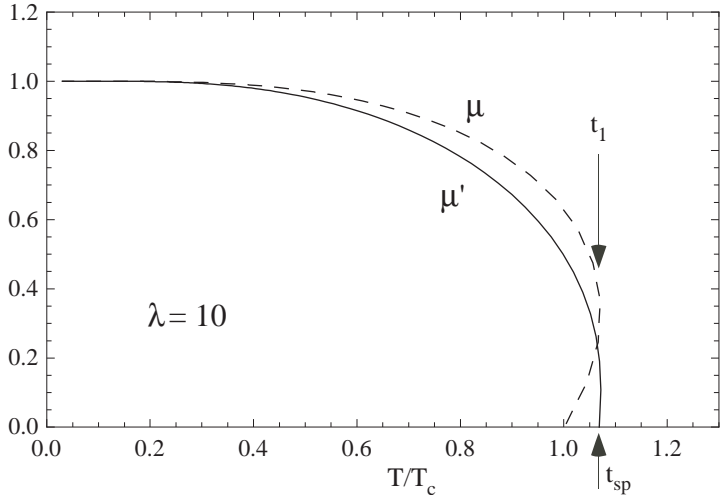}
\caption{Mass shift of the mesonic excitation: solid line ($\mu'$) is the mass of 
the excitation determined by (\ref{mass-shift}) and the dashed line is the solution of 
the gap equation evaluated at $\lambda =$10.  The slopes of two curves diverge 
at the temperature $t_1 = T_1/T_c = 1.0714$ while $\mu'$ vanishes at 
$t_{\rm sp} =T_{\rm sp}/T_c =1.0702$.}

\label{default}
\end{center}
\end{figure}

\section{Concluding remarks}

In this paper, we have developed a kinetic theory for a system of 
interacting quantum fields in the mean field approximation taking into
account the existence of quasi-particle excitations.   We have obtained a set of 
coupled equations of motion, one for the meson condensate in the form 
of non-linear Klein-Gordon equation which contains extra non-linearlity 
due to the particle excitations.  The equations of motion for quasi-particle excitations
are described in terms of the Wigner functions, which reduce to a semiclassical 
Vlasov equation for one-particle distribution function with a modification 
due to the coherent pair-creation and pair-annihilation expressed by the
off-diagonal components in the Wigner function.  These off-diagonal 
components may be eliminated by a suitable Bogoliubov transformation 
of the particle creation and annihilation operators for a uniform, 
time-independent system.    However, they remain non-vanishing in 
general non-uniform, time-dependent systems.

We have shown that in equilibrium these equations are reduced to a
gap equation in the Hartree approximation.  This implies that our kinetic 
equations are natural extension of the Hartree approximation to the 
non-equilibrium situation.  It is well known, however, that in this approximation 
the phase transition becomes first order.   This is a generic feature of 
the mean field approximation~\cite{BG77,ACP93,RM98} which persists 
also for multi-components scalar fields with continuous $O(N)$ symmetry 
for arbitrary finite integer $N$.

In the present work, we studied also the excitations spectrum in the 
system near equilibrium. 
In the high temperature phase there is no collective excitation mode
in the system besides the two continua of the quasi-particle excitations
in the entire space-like energy-momentum region and the time-like
region above the pair creation threshold.   In the low temperature
phase we found that coupling of the meson pole to the quasi-particle
continua give rise to the shift of the meson mass which becomes
zero at the edge of the spinodal instability line of the first order transition. 
The spinodal point appears deep inside the meta-stable region of the
first order transition which may well be an artifact of the mean field 
approximation.           
We found that the coupling to the off-diagonal components of the 
Wigner functions should be properly included to avoid the appearance
of the undamped tachyonic mode.    

In this work we used a single component real scalar field model which 
possesses only discrete symmetry.
It is straightforward to extend the present analysis to models with
continuous symmetry such as the sigma model with $O(N)$ symmetry. 
Basic features of the present analysis is preserved in such extension.   
It is known however that the Goldstone theorem is apparently violated in 
this approximation for a system with continuous symmetry.   We will
show in the forthcoming paper~\cite{MM2}, that the missing 
Nambu-Goldstone mode may be retrieved in the collective excitations
of the system. 

Our coupled kinetic equations may be solved for an arbitrary initial 
conditions.  We plan to study the freezeout dynamics with these 
equations with more realistic interactions.   It would be interesting 
to see in particular how much flow is generated by the acceleration 
by the mean field as the vacuum condensate is restored.  

\vskip 0.5cm

\centerline{\bf Acknowledgment}


We thank Gordon Baym, Hirotsugu Fujii, Tetsuo Hatsuda, Tetsufumi Hirano, 
Osamu Morimatsu and Koichi Ohta for helpful discussions.   
We are indebted to Gordon Baym for correcting a sign error in the original 
manuscript and to Tetsuo Hatsuda and Osama Morimatsu for calling our
attention to some works on the spectral functions at finite temperature 
which prompted us to extend our analyses of the excitation spectrum 
in section 6.  We are also grateful to  Francois Gelis, Kazunori Itakura, 
Yusuke Kato, and Tetsuro Nikuni for their interests in this work. 
This work has been supported by the Grants-in-Aid of the Japanese 
Ministry of Education, Culture, Sports, Science and Technology 
No. 13440067 and 19540269. 

\vskip 20pt

\appendix

\section{Proof of the equivalence of the Gaussian density matrix average
and the mean field approximation.}
Here we present some details of the computation of the equation of motion of the 
Wigner functions and show that with the Gaussian Ansatz for the density matrix
the result is equivalent to what we obtain from the mean-field Hamiltonian (\ref{Hmf}). 

We like to compute the time-derivative of the operator product
$a^\dagger_{\bp_1} a_{\bp_2} $ which appears in the definition of $F( \bp, \bk, t )$ 
with $\bp_1 = \bp + \bk/2$ and $\bp_2 = \bp - \bk/2$. 
\begin{eqnarray}
i \frac{\partial  }{\partial t}  \left( a^\dagger_{\bp_1} a_{\bp_2} \right)
& = & i  {\dot a}^\dagger_{\bp_1} a_{\bp_2} +
i a^\dagger_{\bp_1} {\dot a}_{\bp_2}  \nonumber \\
& = &  [ a^\dagger_{\bp_1}, H ] a_{\bp_2}  +
 a^\dagger_{\bp_1} [ a_{\bp_2}, H ] 
\end{eqnarray}
The commutators $[ a^\dagger_{\bp_1}, H ]$ is decomposed into a sum of
the commutators with $H_i$ ( $i = 1,\cdots,4 $) among which the commutators with
$H_1$ and $H_3$ do not survive the average with the Gaussian density matrix
since they only contain the odd power of the field operators.
We only need to compute the commutators with $H_2$ and $H_4$.

To compute the commutator with $H_2$ it is convenient to rewrite $H_2$ as
\begin{equation}
H_2 = \sum_\bp \omega_{\bp} \left( a^\dagger_\bp a_\bp + \frac{1}{2} \right) 
+ \frac{\lambda}{4} \int d \br \phi_c^2 (\br, t ) \tphi^2 (\br, t) 
\end{equation}
We then find
\begin{eqnarray}
[ a_\bp (t), H_2 ] & = & \omega_{\bp} a_\bp (t) +  \frac{\lambda}{4} \int d \br \phi_c^2 (\br, t ) 
[ a_\bp (t), \tphi^2 (\br, t) ]  \nonumber \\
& = & \omega_{\bp} a_\bp (t) + \frac{\lambda}{2} \int d \br \phi_c^2 (\br, t ) 
\frac{e^{- i \bp \cdot \br}}{\sqrt{2 \omega_{\bp}}} \tphi (\br, t) \\
\left[ a_\bp^\dagger (t), H_2 \right] & = & - \omega_{\bp} a_\bp^\dagger (t) +  
\frac{\lambda}{4} \int d \br \phi_c^2 (\br, t ) 
[ a_\bp^\dagger (t), \tphi^2 (\br, t) ]  \nonumber \\
& = & - \omega_{\bp} a_\bp (t) - \frac{\lambda}{2} \int d \br \phi_c^2 (\br, t ) 
\frac{e^{ i \bp \cdot \br}}{\sqrt{2 \omega_{\bp}}} \tphi (\br, t) 
\end{eqnarray}
where we have used the following formulae:
\begin{eqnarray}
[ a_\bp^\dagger, \tphi^n (\br ,t ) ] 
& = & n \tphi^{n-1}  ( \br , t ) [ a_\bp^\dagger, \tphi (\br ,t ) ] 
= -  n \tphi^{n-1}  ( \br , t ) \frac{e^{ i \bp \cdot \br}}{\sqrt{2 \omega_{\bp}}} \\ 
\left[ a_\bp , \tphi^n ( \br , t ) \right] 
& = & n \phi^{n-1}  (\br ,t ) [ a_\bp, \tphi (\br ,t ) ] 
= n \phi^{n-1}  (\br ,t ) \frac{e^{- i \bp \cdot \br}}{\sqrt{2 \omega_{\bp}}} 
\end{eqnarray}
With these results we obtain 
\begin{eqnarray}
[ a^\dagger_{\bp_1} a_{\bp_2} , H_2 ] 
& = & - ( \omega_{\bp_1} - \omega_{\bp_2} ) a^\dagger_{\bp_1}  a_{\bp_2}  \nonumber \\
& & \quad - \frac{\lambda}{2} \int d \br \phi_c^2 (\br, t ) \tphi (\br, t )
 \frac{e^{ i \bp_1 \cdot \br}}{\sqrt{2 \omega_{\bp_1}}}  a_{\bp_2}   \nonumber \\
& & \quad \qquad  + \frac{\lambda}{2} \int d \br \phi_c^2 (\br, t ) \tphi (\br, t )
\frac{e^{- i \bp_2 \cdot \br}}{\sqrt{2 \omega_{\bp_2}}} a_{\bp_1}^\dagger 
\end{eqnarray} 
The Gaussian average of this equation gives
\begin{eqnarray}
\la [ a^\dagger_{\bp_1} a_{\bp_2} , H_2 ] \ra
& = & - ( \omega_{\bp_1} - \omega_{\bp_2} ) \la a^\dagger_{\bp_1}  a_{\bp_2} \ra \nonumber \\
& & \quad - \frac{\lambda}{2} \int d \br \phi_c^2 (\br, t ) 
 \frac{e^{ i \bp_1 \cdot \br}}{\sqrt{2 \omega_{\bp_1}}} \la \tphi (\br, t ) a_{\bp_2} \ra  \nonumber \\
& & \quad \qquad  + \frac{\lambda}{2} \int d \br \phi_c^2 (\br, t ) 
\frac{e^{- i \bp_2 \cdot \br}}{\sqrt{2 \omega_{\bp_2}}} \la \tphi (\br, t )a_{\bp_1}^\dagger \ra
\end{eqnarray} 

Next we compute the commutators with $H_4$.  We first compute
\begin{eqnarray}
[ a_\bp (t), H_4 ] & = & \frac{\lambda}{3!} \int d \br \tphi^3 (\br, t ) 
\frac{e^{ i \bp \cdot \br}}{\sqrt{2 \omega_{\bp}}} a_\bp (t) \\
\left[ a_\bp^\dagger (t), H_4 \right] & = &
- \frac{\lambda}{3!} \int d \br \tphi^3 (\br, t ) 
\frac{e^{ i \bp \cdot \br}}{\sqrt{2 \omega_{\bp}}} a_\bp (t) 
\end{eqnarray}
We then use these results to obtain
\begin{eqnarray}
[ a^\dagger_{\bp_1} a_{\bp_2} , H_4 ] 
& = & - \frac{\lambda}{3!} \int d \br  \tphi^3 (\br, t )
\frac{e^{ i \bp_1 \cdot \br}}{\sqrt{2 \omega_{\bp_1}}}  a_{\bp_2}  
\nonumber \\
& & \qquad + \frac{\lambda}{3!} \int d \br  \tphi^3 (\br, t )
\frac{e^{- i \bp_2 \cdot \br}}{\sqrt{2 \omega_{\bp_2}}} a_{\bp_1}^\dagger 
\end{eqnarray} 
Taking the average with the Gaussian density matrix, we find
\begin{eqnarray}
\la [ a^\dagger_{\bp_1} a_{\bp_2} , H_4 ] \ra
& = & - \frac{\lambda}{2} \int d \br  \la \tphi^2 (\br, t ) \ra
\frac{e^{ i \bp_1 \cdot \br}}{\sqrt{2 \omega_{\bp_1}}}  \la \tphi (\br, t ) a_{\bp_2}  \ra
\nonumber \\
& & \qquad + \frac{\lambda}{2} \int d \br  \la \tphi^2 (\br, t ) \ra
\frac{e^{- i \bp_2 \cdot \br}}{\sqrt{2 \omega_{\bp_2}}} \la \tphi (\br, t )a_{\bp_1}^\dagger \ra
\end{eqnarray} 

We observe the similarity between the last two terms of the commutator with $H_2$ 
and the commutator with $H_4$ after Gaussian average.  These terms can be 
combined by introducing the self-energy function,
\begin{equation}
\Pi ( \br, t ) = \phi_c^2 (\br, t ) + \la \tphi^2 ( \br, t ) \ra 
= \sum_\bq  \Pi_\bq (t) e^{i \bq \cdot \br }
\end{equation}
Integration over the space coordinate $\br$ give a delta function 
$\delta ( \bq + \bp + \bp_1 )$ and $\delta ( \bq + \bp - \bp_2 )$.
Performing the integral over $\bp$ yields
\begin{eqnarray}
\la [ a^\dagger_{\bp_1} a_{\bp_2} , H] \ra
& = & - ( \omega_{\bp_1} - \omega_{\bp_2} ) \la a^\dagger_{\bp_1}  a_{\bp_2} \ra 
\nonumber \\
& & \quad  - \frac{\lambda}{4} \sum_\bq 
\frac{\Pi_\bq }{\sqrt{\omega_{\bp_1 + \bq}\omega_{\bp_2}}}  
( \la a_{-\bp_1-\bq} a_{\bp_2} \ra + \la a_{\bp_1+ \bq}^\dagger a_{\bp_2} \ra )
\nonumber \\
& &
\qquad  + \frac{\lambda}{4} \sum_\bq 
\frac{\Pi_\bq }{\sqrt{ \omega_{\bp_1} \omega_{p_2 - \bq}}} 
 ( \la a_{\bp_2 - \bq}  a_{\bp_1}^\dagger \ra + 
 \la a_{- \bp_2 + \bq}^\dagger a_{\bp_1}^\dagger \ra )
 \nonumber \\
\end{eqnarray} 
This result coincides with the average of the commutator with the mean field 
Hamiltonian $H_{\rm mf}$ defined by (\ref{Hmf}).
\begin{equation}
\la [ a^\dagger_{\bp_1} a_{\bp_2} , H] \ra = \la [ a^\dagger_{\bp_1} a_{\bp_2} , H_{\rm mf}] \ra
\end{equation}

Commutators of the four bilinear operator products with the mean-field Hamiltonian
are listed below:
\begin{eqnarray}
[ a^\dagger_{\bp_1} a_{\bp_2} , H_{\rm mf} ]  
& = & - ( \omega_{\bp_1} - \omega_{\bp_2} )  a^\dagger_{\bp_1}  a_{\bp_2} 
\nonumber \\
& & \quad  - \frac{1}{2} \sum_\bq 
\frac{\Pi_\bq }{\sqrt{\omega_{\bp_1 + \bq}\omega_{\bp_2}}}  
(  a_{-\bp_1-\bq} a_{\bp_2}  +  a_{\bp_1+ \bq}^\dagger a_{\bp_2}  )
\nonumber \\
& &
\qquad  + \frac{1}{2} \sum_\bq 
\frac{\Pi_\bq }{\sqrt{ \omega_{\bp_1} \omega_{\bp_2 - \bq}}} 
 ( a_{\bp_1}^\dagger  a_{\bp_2 - \bq}   + 
  a_{\bp_1}^\dagger a_{- \bp_2 + \bq}^\dagger  )
  \nonumber \\
  \\
 \left[ a_{\bp_1} a_{\bp_2} , H_{\rm mf} \right] 
& = & ( \omega_{\bp_1} + \omega_{\bp_2} )  a_{\bp_1}  a_{\bp_2} 
\nonumber \\
& & \quad  + \frac{1}{2} \sum_\bq 
\frac{\Pi_\bq }{\sqrt{\omega_{\bp_1 + \bq}\omega_{\bp_2}}}  
(  a_{\bp_1-\bq}  +  a_{-\bp_1+ \bq}^\dagger )  a_{\bp_2}  
\nonumber \\
& &
\qquad  + \frac{1}{2} \sum_\bq 
\frac{\Pi_\bq }{\sqrt{ \omega_{\bp_1} \omega_{\bp_2 - \bq}}} 
 a_{\bp_1} (  a_{\bp_2 - \bq}   + a_{- \bp_2 + \bq}^\dagger  )
 \nonumber \\
 \\
\left[ a^\dagger_{\bp_1} a^\dagger_{\bp_2} , H_{\rm mf} \right]  
& = & - ( \omega_{\bp_1} + \omega_{\bp_2} )  a^\dagger_{\bp_1}  a_{\bp_2} 
\nonumber \\
& & \quad  - \frac{1}{2} \sum_\bq 
\frac{\Pi_\bq }{\sqrt{\omega_{\bp_1 + \bq}\omega_{\bp_2}}}  
(  a_{-\bp_1-\bq} a_{\bp_2}  +  a_{\bp_1+ \bq}^\dagger a_{\bp_2}  )
\nonumber \\
& &
\qquad  - \frac{1}{2} \sum_\bq 
\frac{\Pi_\bq }{\sqrt{ \omega_{\bp_1} \omega_{\bp_2 - \bq}}} 
 (  a_{\bp_2 - \bq}  a_{\bp_1}^\dagger  + 
  a_{- \bp_2 + \bq}^\dagger a_{\bp_1}^\dagger )
 \nonumber \\
 \\
 \left[ a_{\bp_1} a^\dagger_{\bp_2} , H_{\rm mf} \right]  
& = & ( \omega_{\bp_1} - \omega_{\bp_2} )  a_{\bp_1}  a^\dagger_{\bp_2} 
\nonumber \\
& & \quad  + \frac{1}{2} \sum_\bq 
\frac{\Pi_\bq }{\sqrt{\omega_{\bp_1 + \bq}\omega_{\bp_2}}}  
(  a_{\bp_1-\bq} +  a_{- \bp_1+ \bq}^\dagger ) a^\dagger_{\bp_2}  
\nonumber \\
& &
\qquad  - \frac{1}{2} \sum_\bq 
\frac{\Pi_\bq }{\sqrt{ \omega_{\bp_1} \omega_{\bp_2 - \bq}}} 
 a_{\bp_1}^\dagger   (  a_{\bp_2 - \bq}  + a_{- \bp_2 + \bq}^\dagger )
 \nonumber \\
\end{eqnarray} 





\begin{thebibliography}{}



\def\AP#1{{Ann. Phys. }{\bf #1}}
\def\PRP#1{{Phys. Rep. }{\bf #1}}
\def\APP#1{{Act. Phys. Pol. }{\bf #1}}
\def\PTP#1{{Prog. Theor. Phys. }{\bf #1}}
\def\PTPS#1{{Prog. Theor. Phys. Supplement}{\bf #1}}
\def\PR#1{{Phys. Rev. }{\bf #1}}
\def\PRD#1{{Phys. Rev. }{\bf D#1}}
\def\PRC#1{{Phys. Rev. }{\bf C#1}}
\def\PRA#1{{Phys. Rev. }{\bf A#1}}
\def\PRL#1{{Phys. Rev. Lett. }{\bf #1}}
\def\PL#1{{Phys. Lett. }{\bf #1}}
\def\RMP#1{{Rev. Mod. Phys.}{\bf #1}}
\def\NP#1{{Nucl. Phys. }{\bf #1}}
\def\ZP#1{{Z. Phys. }{\bf #1}}
\def\NC#1{{Nuovo Cimento }{\bf #1}}
\def\SJNP#1{{Sov. J. Nucl. Phys. }{\bf #1}}
\def\EPJA#1{{Eur. Phys. J. }{\bf A#1}}
\def\CF#1{{Coll. Phen. }{\bf A#1}}


\bibitem{Fer50} E. Fermi, \PTP{5}, 570 (1950); L. D. Landau, Izv. Akad. Nauk SSSR, {\bf 17}, 51(1953)

\bibitem{Bjo83} J. D. Bjorken, {\it Phys. Rev.} {\bf D27}, 865 (1983) ; 
G. Baym, B. L. Friman, J.-P. Blaizot, M. Soyeur, W. Czy\'z, {\it Nucl. Phys.} {\bf A407}, 541 (1983);
K. Kajantie and L. McLerran, \NP{B214}, 261 (1983) ; 
M. Gyulassy, T. Matsui, {\it Phys. Rev.} {\bf D29}, 419 (1984) 

\bibitem{RHICWP05} B. Back {\it et al. } (PHOBOS Collaboration), \NP{A757}, 28 (2005) ;
J. Adama {\it et al.} (STAR Collaboration), \NP{A757} ,102 (2005); K. Adcox, {\it et al.} (PHENIX Collaboration). \NP{A757}, 184 (2005)

\bibitem{Flow}  
P.F. Kolb, P. Huovinen, U. Heinz, H. Heiselberg, Phys. Lett. B500, 232 (2001); T. Hirano, \PRC{65} 
011901 (2001) 

\bibitem{Bay84} Some early attempts to describe the early thermalization processes in terms 
of kinetic theory can be found, for example, in G. Baym, \PL{B138}, 18 (1984); 
K. Kajantie and T. Matsui, \PL{B164}, 373 (1985). 

\bibitem{BS99} P. Braun-Munzinger and J. Stachel, \PL{B465}, 15 (1999)

\bibitem{HBT} M. Lisa, S. Pratt, R. Soltz, U. Wiedemann,  Ann. Rev. Nucl. Part. Sci. {\bf 55}, 357
 (2005) 

\bibitem{NJL61} Y. Nambu, and G. Jona-Lasinio, \PR{122}, 345 (1961); \PR{124}, 246 (1961); 
T. Hatsuda, and T. Kunihiro; \PRP{247}, 221 (1994). 

\bibitem{Bjo87} J. D. Bjorken,  Int. J. Mod. Phys. A7, 4189  (1987)

\bibitem{RW93} K. Rajagopal, F. Wilczek, \NP{B404}, 577 (1993); S. Gavin, A. Gocksch, and R. D. Pisarski, \PRL{72} 2443 (1994); M. Asakawa. Z. Huang and X.-N. Wang, \PRL{74}, 3126 (1995)  

\bibitem{TVM99} Y.~Tsue, D.~Vautherin and T.~Matsui, \PTP{102}, 313 (1999);
Y.~Tsue, D.~Vautherin and T.~Matsui, \PRD{61}, 076006 (2000).

\bibitem{KV89} A. Kerman, and D. Vautherin, \AP{192} , 408 (1988)

\bibitem{EJP88} O. \'Eboli, R. Jackiw, and S.-Y. Pi, \PRD{37} , 3557 (1988)

\bibitem{PS02} C. J. Pethick, and H. Smith, {\it Bose-Einstein Condensation in Dilute Gases} 
(Cambridge Univ. Press, 2002)

\bibitem{GP61} E. P.~Gross, Nuovo Cimento, {\bf 20}, 454 (1961); J. Math. Phys. {\bf 4},
195 (1963); L.~P.~Pitaevskii, Zh. Eks. Theor. Fiz. {\bf 40}, 646 (1961)
[Sov. Phys. JETP {\bf 13}, 451 (1961)].

\bibitem{ZNG99} E. Zaremba, T. Nikuni, and A. Griffin, {\it J. Low Temp. Phys.} {\bf 116}, 277 (1999)

\bibitem{ITG99} M. Imamovi\'c-Tomasovi\'c,  and A. Griffin, \PRA{60} , 494 (1999)

\bibitem{KB62} L. Kadanoff, and G. Baym, {\it Quantum Statistical Mechanics} 
(W. A. Benjamin, Inc., 1962) 

\bibitem{CH88} For example see, E. Calzetta, B. L. Hu, \PRD{37}, 2878 (1988); 
A. H. Mueller, and D. T. Son, \PL{B582}, 279 (2004)

\bibitem{ALMY05} P. Arnold, J. Lenaghan, G. Moore, and L. Yaffe, \PRL{94}, 072302 (2005)

\bibitem{ABM06} Y. Asakawa, S. A. Bass, and B. M\"uller, \PRL{96}, 252301 (2006)

\bibitem{MM2} M. Matsuo, and T. Matsui, in preparation.

\bibitem{FW71} A. Fetter, and J. D. Walecka, {\it Quantum Theory of Many-Particle Systems}
(McGraw-Hill, Co., 1971) 

\bibitem{Gor58} L. P. Gorkov,  Sov. Phys. JETP {\bf 7}, 505 (1958)

\bibitem{Nam60} Y. Nambu, \PR{117}, 648 (1960)

\bibitem{DJ74} L. Dolan and R. Jackiw, \PRD{9}, 3320 (1974)

\bibitem{BG77} G. Baym, and G. Grinstein, \PRD{15}, 2897 (1977)

\bibitem{Lan57} L. D. Landau, Sov. Phys. JETP {\bf 3}, 920 (1957); {\it ibid. } {\bf 5},
101 (1957)

\bibitem{BP91} G. Baym and C. Pethick, {\it Landau Fermi-Liquid Theory: concepts and applications}
(John Wiley \& Sons Inc., 1991)

\bibitem{HK85} T. Hatsuda and T. Kunihiro, \PRL{55}, 158 (1985)  

\bibitem{BC76} G. Baym and S. A. Chin, \NP{A262}, 527 (1976)

\bibitem{Mat81} T. Matsui, \NP{A370}, 365 (1981) 

\bibitem{Wal74} J. D. Walecka, \AP{83}, 491 (1974); 
B. D. Serot and J. D. Walecka, {\it Advances in Nuclear Physics} {\bf 16}, 
eds. J. W. Negele and E. Vogt, Plenum Press, New York (1986)

\bibitem{RM98} H.-S. Roh and T. Matsui, \EPJA{1}, 205 (1998)

\bibitem{CH98} S. Chiku, and T. Hatsuda, \PRD{58}, 076001 (1998)

\bibitem{Ch00} S. Chiku, \PTP{104}, 1129 (2000)

\bibitem{ACP93} G. Amelino-Camelia and S.-Y. Pi, \PRD{47}, 2356 (1993)

\bibitem{CJT74} J. M. Cornwall, R. Jackiw, and E. Tomboulis, \PRD{10}, 2428 (1974)

\bibitem{LP} E. M. Lifshitz and L. P. Pitaevskii, {\it Physical Kinetics} (Pergamon,1981) ch. 3.

\bibitem{BI02}  See, for example, J.-P. Blaizot and E. Iancu, \PRP{359}, 355 (2002)

\bibitem{Bog47} N. N. Bogoliubov, J. Phys. (USSR) {\bf 11}, 23 (1947) 

\bibitem{NS84} A. J. Niemi, and G. W. Semenoff, \AP{152}, 105 (1984); \NP{B230}, 181 (1984)


\bibitem{Wel83} H. A. Weldon, \PRD{28}, 2007 (1983)

\bibitem{PSS02} A.~Patkos, Z.~Szep and P.~Szepfalusy,
\PL{B537}, 77 (2002);  \PRD{66}, 116004 (2002); {\it ibd} 047701 (2003)

\bibitem{HMN03} Y. Hidaka, O. Morimatsu, T. Nishikawa, \PRD{67}, 056004 (2003)

\bibitem{VH54} L. van Hove, \PR{95}, 249 (1954)

\bibitem{PN66} D. Pines and P. Nosi\'eres, {\it The Theory of Quantum Liquids I: Normal
Fermi Liquids}  (Benjamin, Inc., 1966)

\bibitem{NP90} P. Nosi\'eres and D. Pines,  {\it The Theory of Quantum Liquids II: Superfluid 
Bose Liquids}  (Addison-Wesley, 1990)



\end{thebibliography}
\end{document}